\documentclass[usenatbib]{mn2e}

\usepackage[english]{babel}
\usepackage[babel]{csquotes}
\usepackage{courier}
\usepackage{setspace}
\usepackage{fancyhdr}
\usepackage{graphicx}
\usepackage{booktabs}
\usepackage{amsmath}
\usepackage{tensor}
\usepackage{placeins}
\usepackage{natbibmnfix}
\usepackage{epsfig}
\usepackage{aas_macros}
\usepackage{color}

\def\gtsima{$\; \buildrel > \over \sim \;$} 
\def\ltsima{$\; \buildrel < \over \sim \;$} 
\def\gsim{\lower.5ex\hbox{\gtsima}} 
\def\lsim{\lower.5ex\hbox{\ltsima}} 
\def\simgt{\lower.5ex\hbox{\gtsima}} 
\def\simlt{\lower.5ex\hbox{\ltsima}} 

\def\Lya{Ly$\alpha$~}
\def\Lyb{Ly$\beta$~}
\def\lya { {Ly\alpha} }
\def\HI{\hbox{H~$\scriptstyle\rm I\ $}}

\pdfoutput=1

\title[Ly$\alpha$ intensity mapping]
 {Probing high-redshift galaxies with \Lya  intensity mapping}
\author[P. Comaschi \& A. Ferrara]
  {P.~Comaschi$^1$\thanks{Email: paolo.comaschi@sns.it}
  and A.~Ferrara$^{1,2}$\\
  $^1$Scuola Normale Superiore, Piazza dei Cavalieri 7, 1-56126 Pisa, Italy\\
  $^2$Kavli IPMU, The University of Tokyo, 5-1-5 Kashiwanoha, Kashiwa 277-8583, Japan}
\date{}

\def\LaTeX{L\kern-.36em\raise.3ex\hbox{a}\kern-.15em
    T\kern-.1667em\lower.7ex\hbox{E}\kern-.125emX}

\begin{document}
	
	\maketitle

\begin{abstract}
We present a study of  the cosmological \Lya emission signal at $z > 4$. Our goal is to predict the power spectrum of the spatial fluctuations that could be observed by an intensity mapping survey. The model uses the latest data from the HST legacy fields and the abundance matching technique to associate UV emission and dust properties with the halos, computing the emission from the interstellar medium (ISM) of galaxies and the intergalactic medium (IGM), including the effects of reionization, self-consistently. 
The \Lya intensity from the diffuse IGM emission is 1.3 (2.0) times more intense than the ISM emission at $z = 4(7)$; both components are fair tracers of the star-forming galaxy distribution. However the power spectrum is dominated by ISM emission on small scales  ($k > 0.01 h{\rm Mpc}^{-1}$) with shot noise being significant only above $k = 1 h{\rm Mpc}^{-1}$. At very lange scales ($k < 0.01h{\rm Mpc}^{-1}$) diffuse IGM emission becomes important. The comoving \Lya luminosity density from IGM and galaxies, $\dot \rho_\lya^{\rm IGM}  = 8.73(6.51) \times 10^{40} {\rm erg~}{\rm s}^{-1}{\rm Mpc}^{-3}$ and $\dot \rho_\lya^{\rm ISM}  = 6.62(3.21) \times 10^{40} {\rm erg~}{\rm s}^{-1}{\rm Mpc}^{-3}$ at $z = 4(7)$, is consistent with recent SDSS determinations.
We predict a power  $k^3 P^\lya(k, z)/2\pi^2 = 9.76\times 10^{-4}(2.09\times 10^{-5}){\rm nW}^2{\rm m}^{-4}{\rm sr}^{-2}$ at $z = 4(7)$ for $k = 0.1 h {\rm Mpc}^{-1}$. 
\end{abstract}

\begin{keywords}
 cosmology: observations - intergalactic and interstellar medium - intensity mapping - large-scale structure of universe
\end{keywords}

\section{Introduction}
\label{sec:intro}
According to the most popular cosmological framework, the $\Lambda$CDM model, the astonishing diversity present in the local universe originated from tiny density fluctuations in a homogeneous hot plasma. During the Hubble expansion, the dark matter (DM) in the most overdense regions started to collapse through gravitational instabilities, forming virialized halos where baryons could cool and form stars. This process gave birth to primordial galaxies and, with billions of years, every structure we can observe today. The impact of these early objects on the subsequent cosmic evolution, mediated by a number of feedback processes, has been dramatic; in addition, their radiative energy input powered the last phase transition in the universe during the Epoch of Reionization (EoR, \citep{2001PhR...349..125B}).  

It is quite surprising that, for fifty years, we have been able to observe directly both the beginning, through the CMB, and the ending of this process. However the study of the ancient universe proved to be extremely challenging and only in the last decade we have been able to explore the EoR latest stages.       

One of the main challenges is that the first galaxies are very faint and hard to detect even with our most powerful telescopes. Up to date, the deepest ``drop-out'' photometric surveys carried out by the Hubble Space Telescope (HST), could detect $\sim 700$ galaxies at $z \geq 8 $ \citep{2014ApJ...793..115B}. Moreover, such sources are the most massive outliers of the galaxy population and therefore bear little information about the sources producing the bulk of the ionizing radiation \citep{2011MNRAS.414..847S}. This difficulty will likely not be overcome by even the next generation of deep surveys, such as the JWST one. 

For this reason, a different strategy must be designed, which implies to forego the detection of individual sources and probe directly their large scale distribution. This idea can be implemented through the Intensity Mapping (IM) of selected emission lines \citep{2010JCAP...11..016V, 2011JCAP...08..010V}. Each point in space is identified by an angular coordinate and a redshift, that can be measured knowing the frequency of the emitted and detected photons. If the relevant foregrounds can be removed, one can measure the distribution of the cumulative galaxy emission in coarse voxels. For example, if the galaxy-galaxy correlation length is 1 Mpc and we are interested in structures larger than 10 Mpc, we can probe the sky in 10 Mpc voxels, corresponding to a spectral (angular) resolution $\lambda/\delta \lambda \approx 300$ ($\theta \approx 4'$) at $z=7$, by grouping together the signal of $\sim 1000$ galaxies. 

Several emission lines have been proposed for this kind of survey: 21cm \citep{2006PhR...433..181F}, CO \citep{2011ApJ...741...70L, 2008A&A...489..489R, 2014MNRAS.443.3506B}, ${\rm H}_{2}$ \citep{2013ApJ...768..130G}, [CII] \citep{2012ApJ...745...49G, 2014arXiv1410.4808S, bin}, HeII \citep{2015arXiv150103177V} and, of course, Ly$\alpha$ \citep{2013ApJ...763..132S, 2014ApJ...786..111P}. Noticeably, each line probes different physical processes and, therefore, bears complementary information on the sources. 

The \Lya line most likely represents the optimal feature for a IM experiment. Historically, it has constantly been used for high-$z$ galaxy searches as it is the most luminous UV line. Observers have undertaken deep surveys to detect high redshift \Lya emitters (LAE) \citep{2008ApJS..176..301O, 2010ApJ...723..869O, 2015arXiv150207355M}. These searches are affected by the same problems mentioned above in the context of drop-out searches, with the additional complication arising from the fact that intergalactic \HI can scatter the bulk of \Lya photons out of the line of sight, making systematic detections of LAE during the EoR very challenging \citep{2011MNRAS.410..830D}. IM can overcome these problems, thanks to its sensitivity to even diffuse emission from the IGM. Therefore \Lya IM seems a very promising tool to study the properties of early, faint and distributed EoR sources.

\cite{2013ApJ...763..132S} studied this problem with seminumerical tools, focusing on the EoR emission at $z=7$. In a following work, \cite{2014ApJ...785...72G} tackled the problem of line confusion, proposing the masking of contaminated pixels as a cleaning technique. In these works it is found that recombinations in the interstellar medium (ISM) of galaxies largely dominate the \Lya intensity and power spectrum (PS). 
Finally, \cite{2014ApJ...786..111P} developed a simple analytical model to study the evolution of the \Lya power spectrum at $z > 2$: their results are qualitatively different from \cite{2013ApJ...763..132S}, in that they conclude that diffuse IGM emission is the dominant source. 
Recently \cite{2015arXiv150404088C} attempted to observe the large scale clustering of \Lya emission at $z=2-3$ by cross-correlating the residuals in the SDSS spectra with QSOs. They claim a detection of a mean \Lya surface brightness $\simgt 10$ times more intense than the one inferred from LAE surveys, but still compatible with the unobscured \Lya emission expected from LBGs.  

In this work we develop a detailed analytical model for the \Lya emission and PS. We will use the latest data from the high redshift LBG in the HST legacy fields to compute the emission of ionizing and UV radiation and dust obscuration. Then we will model the physics of the ISM and of the IGM to obtain a data-constrained reionization history and expected \Lya intensity. 
Our main aim is to extract the physical information encoded in a \Lya intensity map, and assess whether it can be used to probe the earliest generation of galaxies, their ISM and also the structure of the IGM during the EoR.

The paper is organized as follows: in \S \ref{sec:starform} we introduce our model for star formation and in \S \ref{ssec:reion} we compute the reionization history accordingly; in \S \ref{sec:meanint} we review the physics involved in Ly$\alpha$ emission and we use it to predict the mean Ly$\alpha$ intensity from the different processes involved. Finally, in \S \ref{sec:fluct} we derive the power spectrum of the \Lya fluctuations and discuss the results. We assume a flat $\Lambda$CDM cosmology compatible with the latest Planck results ($h =0.677$, $\Omega_m = 0.31$, $\Omega_b = 0.049$, $\Omega_\Lambda = 1-\Omega_m$, $n = 0.97$, $\sigma_8 = 0.82$, \citet[]{2015arXiv150201589P}).

\section{High redshift galaxies}
\label{sec:starform}
The first step towards the computation of the \Lya intensity fluctuations is to accurately model the UV emissivity of high-redshift galaxies and their contribution to IGM reionization. This is because \Lya emission mostly arises as a result of recombinations in gas ionized by young, massive stars. Ionizing photons are produced also by  Active Galactic Nuclei; however, as there is a general consensus that reionization is largely driven by stars (but see \citet{2015arXiv150202562G} and \citet{2015MNRAS.448.3167W}), we will neglect in the following the presence of AGNs. In the following we model the UV emissivity of galaxies using an empirical approach based on the most recent UV Luminosity Functions measurements.	

\subsection{UV emissivity}
\label{ssec:sfrd}
As \Lya emission is intimately related to massive stars which have a lifetime (few Myr) short compared to the Hubble time, it effectively depends only on the instantaneous Star Formation Rate (SFR), $\psi$,  of the galaxy. We seek a relation between the halo mass, $M$, and the UV luminosity at 1600\AA, $L_{1600}$. To this aim we use the abundance matching technique \citep{bin, 2009ApJ...696..620C, 2010ApJ...717..379B, 2004MNRAS.353..189V}, which makes the implicit assumption that such relation is monotonic.  The UV luminosity function has been measured recently at $z>8$ using the HST legacy fields \citep{2014ApJ...793..115B, 2014arXiv1403.4295B} and is well fitted by a Schechter parametrization \citep{1976ApJ...203..297S}:
\begin{equation}
\frac{dn({M_{1600}}, z)}{dM_{1600}}= 0.4 \ln(10) \phi^\star x^{1+\alpha} e^{-x},
\end{equation}
where $x = 10^{0.4(M_{1600}^\star - M_{1600} ) }$ and $M_{1600}$ is the absolute dust attenuated AB magnitude at $1600$\AA.
\cite{2014arXiv1403.4295B} computed the redshift evolution of the three Schechter parameters $(\phi^\star, M_{1600}^\star, \alpha)$ with a linear fit of the observational data from $z=4-8$ dropout galaxies, finding significant evolution in $\phi_\star$ and $\alpha$:
\begin{gather}
\label{mstar}
M_{1600}^\star = (-20.95 \pm 0.10) + (0.01 \pm 0.06) (z-6) \\
\phi^\star = (0.47_{-0.10}^{+0.11}) 10^{(-0.27 \pm 0.05)(z-6)} 10^{-3} {\rm Mpc}^{-3} \\
\alpha = (-1.87 \pm 0.05) + (-0.10 \pm 0.03)(z-6).
\end{gather}
We extrapolated these fits to higher redshifts. With this information and the mass function $dn/dM$ \citep{1974ApJ...187..425P, 1999MNRAS.308..119S, 2001MNRAS.323....1S} we can derive a $M_{1600} = M_{1600}(M, z)$ by forcing the equality between the number density of the most luminous galaxies and most massive halos:
\begin{equation}
    \int_M^{+\infty} \frac{dn({M'}, z)}{dM'} d{M'} = \int_{M_{1600}(M, z)}^{+\infty} \frac{dn({\overline M_{1600}}, z)}{d\overline M_{1600}} d{\overline M_{1600}}. 
\end{equation}
  
Since the SFR is correlated with the unobscured UV magnitude $M'_{1600}$, we have to correct for dust. We will use the standard correction using the spectral slope $\beta$ ($f_\lambda \propto \lambda^\beta$) fitted by \cite{2014arXiv1403.4295B}
\begin{gather}
  \label{attenuation}
  \beta(M_{1600}, z) = \beta_{-19.5}(z)  + \frac{d\beta(z)}{dM_{1600} }(M_{1600} + 19.5) \\
  \beta_{-19.5}(z) = -1.97 -0.06(z-6) \\
  \frac{d\beta(z)}{dM_{1600}} = -0.18-0.03(z-6).
\end{gather}

  Then we have \citep{1999ApJ...521...64M}
\begin{gather}
    M'_{1600} = M_{1600} - A_{1600} \\
    A_{1600} = 4.43 + 1.99\beta .
\end{gather}

\begin{figure}
\vspace{+0\baselineskip}
{
\includegraphics[width=0.45\textwidth]{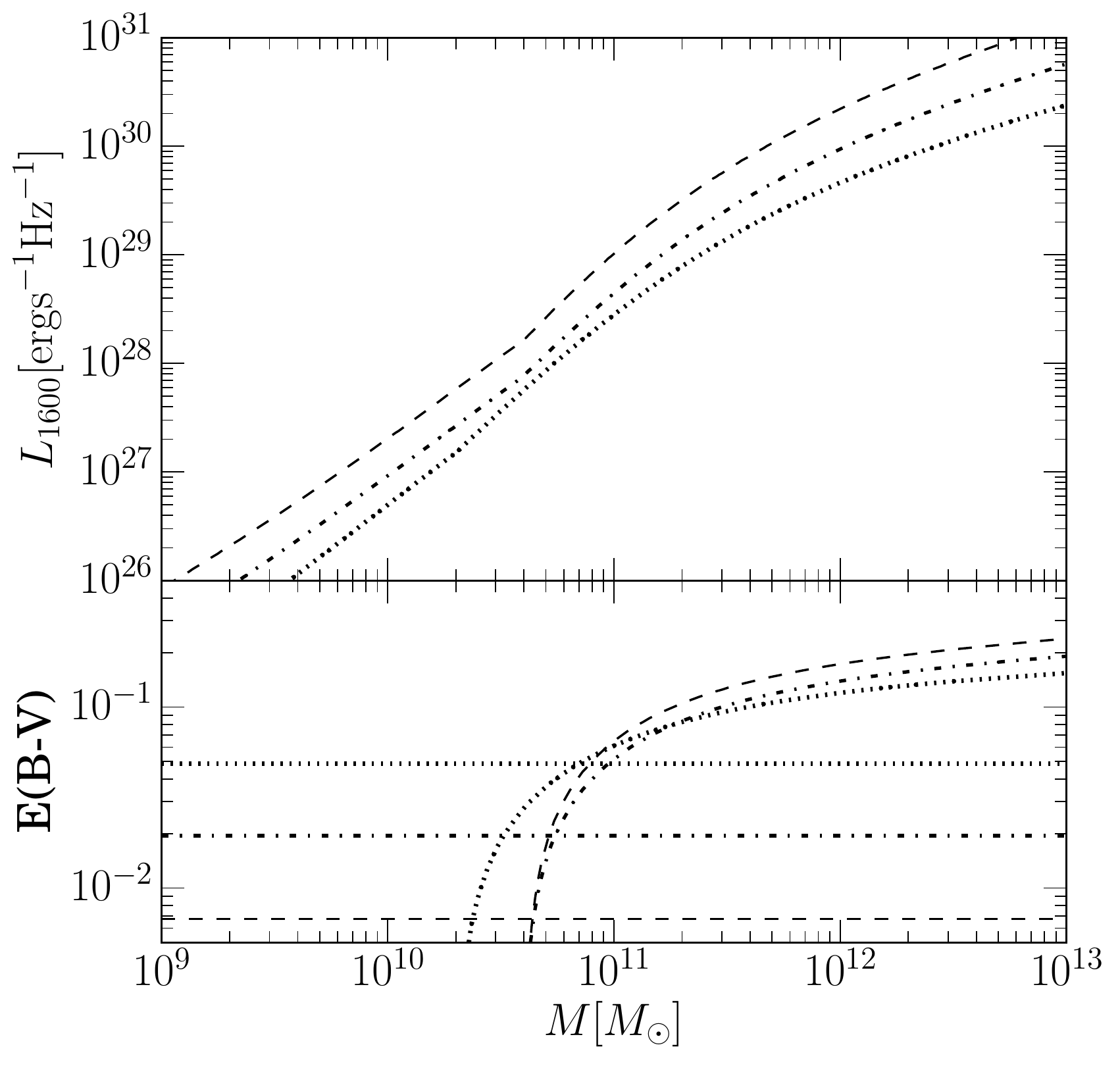}
}
\caption{{\bf Top panel:} relation between UV luminosity at $1600$\AA ~and halo mass, $L_{1600}(M, z)$, obtained from abundance matching and the HST LBGs from \protect\cite{2014arXiv1403.4295B}. The dotted, dot-dashed and dashed lines correspond to $z = $ 4, 6, and 8. {\bf Bottom panel:} dust reddening, E(B-V), from eq. \eqref{attenuation} and \eqref{ebv}. We also show the mean $\langle {\rm E(B-V)} \rangle_{L_\lya}$ (horizontal lines) weighted with \Lya luminosity (eq. \eqref{lyalum}). }
\vspace{-1\baselineskip}
\label{fig:sfr}
\end{figure}

The top panel of Fig. \ref{fig:sfr} shows the resulting $L_{1600}(M, z)$, the intrinsic spectral luminosity density at 1600\AA, at $z =~4$, 6 and 8; the bottom panel shows the reddening E(B-V) (see eq. \eqref{ebv}), along with the average weighted with the \Lya luminosity (eq. \eqref{lyalum}). At higher redshift galaxies are more efficient in forming stars and less dusty. 

Using the stellar synthesis code \texttt{starburst99}\footnote{http://www.stsci.edu/science/starburst99/docs/default.htm} \citep{1999ApJS..123....3L, 2005ApJ...621..695V, 2014ApJS..212...14L} we can compute the intrinsic emission of ionizing photons given the dust corrected UV luminosity: 
\begin{gather}
  L_{1600} = L_{1600}^{1M_\odot/{\rm yr}} \times \psi \\
  \dot N_{\rm 912} = \dot N_{\rm 912}^{1M_\odot/{\rm yr}} \times \psi, 
\end{gather}
finally yielding 
\begin{gather}
\dot N_{\rm 912} = \dot N_{\rm 912}^{1M_\odot/{\rm yr}} \frac{L_{1600}}{L_{1600}^{1M_\odot/{\rm yr}}},
\end{gather}
where $\dot N_{912}$ is the rate of emission of ionizing photons. This treatment introduces two new parameters, such as the age of the stellar population, $t_{\rm age}$ and the metallicity $Z$. 

\begin{figure}
\vspace{+0\baselineskip}
{
\includegraphics[width=0.45\textwidth]{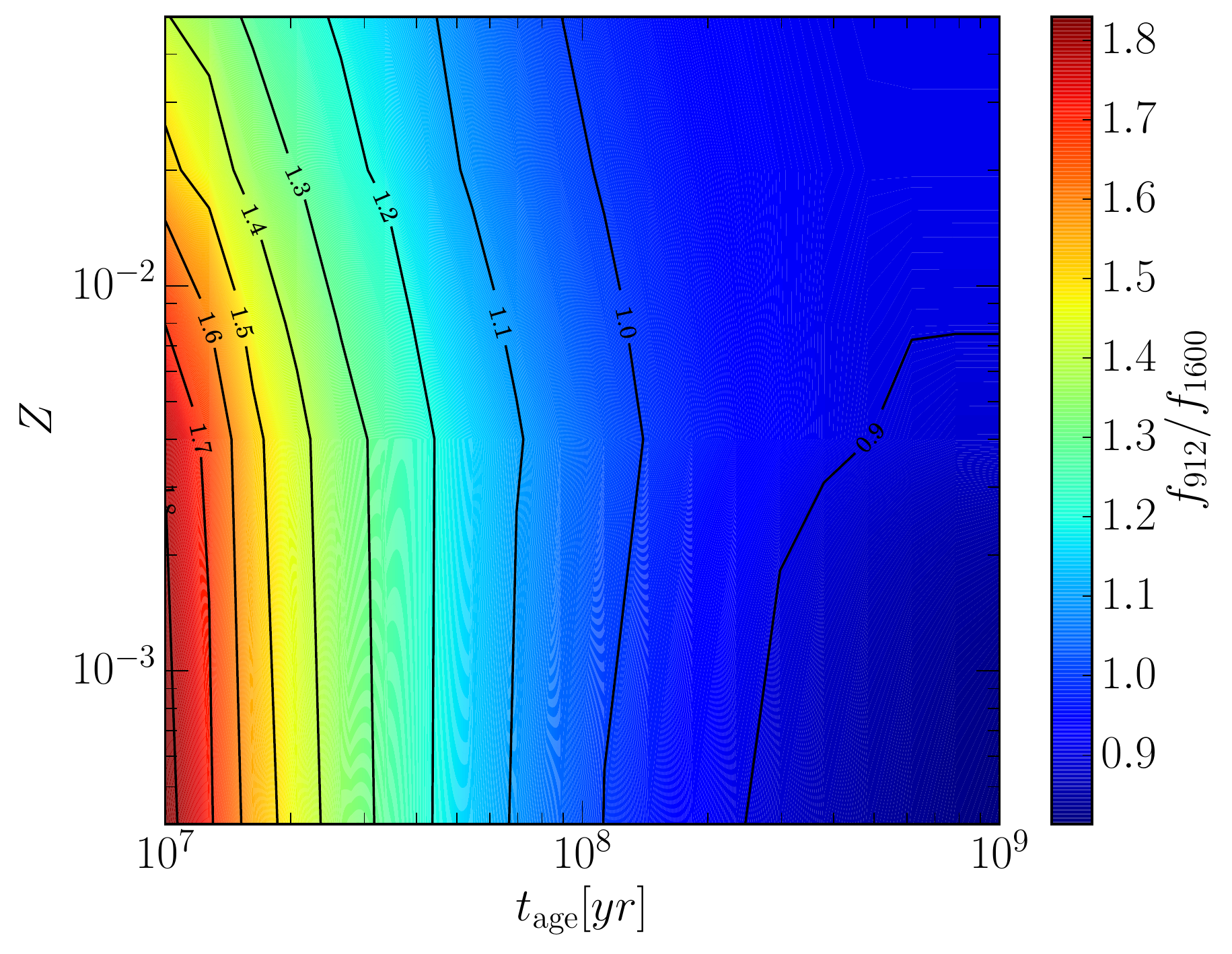}
}
\caption{$L_{912}/L_{1600}$ as function of the stellar age $t_{\rm age}$ and metallicity $Z$. A constant SFR and a Salpeter IMF between 0.1 and 100 $M_\odot$ has been assumed.}
\vspace{-1\baselineskip}
\label{fig:sb99contour}
\end{figure}

Fig. \ref{fig:sb99contour} shows how the ratio $L_{912}/L_{1600}$ depends on these parameters for a galaxy with constant SFR. If we consider $t_{\rm age}\sim 10^8$ yr, i.e. about 10\% of the Hubble time ($\tau_H$) during the EoR, the emission of ionizing photons depends weakly on $Z$ and a factor 2 in $t_{\rm age}$ introduces a order 10\% variation. Therefore in the following we will simply consider $t_{\rm age}(z) = 0.1 \tau_H(z)$, $Z = 0.0004$ and a Salpeter Initial Mass Function (IMF, \cite{1955ApJ...121..161S}) between 0.1 and $100 M_\odot$, because further complications (such as the introduction of a mass-age relation or different star formation histories) would be ill-constrained and would not make our conclusions more solid. At $z > 10$ we imposed $t_{\rm age}(z) = 5 \times 10^7$yr, in order to avoid strong variations in $N_{912}$ due to extremely young stellar ages. 

We need to set a minimum mass $M_{\rm min}$ for star-forming halos; this choice is tricky, as feedback effects on the suppression of star formation in small halos are not fully understood. Although the value of  $M_{\rm min}$ is irrelevant at $z < 8$, it becomes fairly important at high redshift (the difference between models with $M_\mathrm{min}=10^8 M_\odot$ and $M_\mathrm{min}=10^9 M_\odot$ is about 50\% at $z=12$), where the bulk of star formation takes place in small halos \citep{Yue14}.

We set
\begin{equation}
M_\mathrm{min} = 10^8 M_\odot h^{-1}  \left(\frac{\Omega_m \Delta_c}{\Omega_m(z) 18\pi^2}\right)^{1/2}  \left(\frac{1+z}{10}\right)^{-3/2}
\end{equation}
as our fiducial choice\footnote{Note that this choice leads to a star formation rate density (SFRD), $\dot \rho_\star$, higher than typically found by observations (e.g. \cite{2014arXiv1403.4295B}) because we are extrapolating the luminosity function to magnitudes well below the observational limits.}; such minimum mass corresponds to halos with virial temperature of $10^4 K$, the minimum temperature necessary for hydrogen atomic cooling to become efficient.

\subsubsection{Population III stars}
\label{sssec:popIII}
Metal-free (usually defined as PopIII) stars, depending on their IMF, can produce more UV photons per baryon with respect to standard PopII stars.  Since the \Lya intensity is proportional to the integrated emission of these photons, this could ease the detection from high redshift.

In order to estimate the PopIII emission, we used stellar models from \cite{2002A&A...382...28S}. Under the constant SFR hypothesis we can compute the number of photons emitted per baryon \citep{2014ApJ...786..111P}:
\begin{equation}
N = \frac{m_b}{m_\star} \int dm f(m) Q(m) \tau(m),
\end{equation} 
where: $m_b$ is the mean mass of the baryons; $f(m)$ is the IMF; $m_\star$ is the first moment of $f(m)$; $Q(m)$ is the emission rate of photons of a star of mass $m$ while $\tau(m)$ is its the lifetime. By comparing the emission efficiency of PopII and PopIII stars it is possible to simply understand how the \Lya intensity scales with the different stellar populations.

\cite{2002A&A...382...28S} computed $Q(m)$ and $\tau(m)$ for PopII ($Z = Z_\odot/50$), for PopIII stars and for ionizing or Lyman-Werner (LW, $E \in [11.2, 13.6]$eV)  photons. In order to roughly estimate the emission rate of $E=[10.2, 13.6]$eV photons, we added a factor 1.4 to $Q_{\rm LW}$ (but this extrapolation does not affect the results of this Section). We considered also different IMFs, using the PopII and PopIII IMF from \citet{2003PASP..115..763C}. 

Table \ref{tab:popIIIII} shows the ratio of the number of ionizing and continuum photons emitted per baryon by PopII and PopIII stars. For all the combinations considered, the number of photons emitted per baryon is of the same order of magnitude: therefore it is unlikely that PopIII stars could be exploited for the detection of the \Lya signal at high redshifts. For this reason in the following we will neglect the possible presence of PopIII stars in our sources.

\begin{table}
\centering
\begin{tabular}{cccc}
\toprule
 $N_{912}^{\rm III} / N_{912}^{\rm II}$  & Salp.  & ChabII & ChabIII \\
\midrule
Salp. & 1.35 & 3.11 & 1.93 \\
ChabII & 0.75 & 1.72 & 1.07 \\
ChabIII &  1.20 & 2.76 & 1.71 \\
\bottomrule
\end{tabular}
\begin{tabular}{cccc}
\toprule
 $N_{\rm cont}^{\rm III} / N_{\rm cont}^{\rm II}$  & Salp.  & ChabII & ChabIII \\
\midrule
Salp. & 1.14 & 2.80 & 1.53 \\
ChabII & 0.57 & 1.39 & 0.76 \\
ChabIII &  1.06 & 2.61 & 1.43 \\
\bottomrule
\end{tabular}
\caption{The tables show the ratio of the emission rate of ionizing (top) and E=[10.2, 13.6] UV (bottom) photons between PopII and PopIII star. Each row(column) shows a different IMF for PopIII (PopII) stars. Salp. stands for a Salpeter IMF \protect\citep{1955ApJ...121..161S} between 1 and 100 $M_\odot$; ChabII (ChabIII) is the PopII (PopIII) IMF from \protect\citep{2003PASP..115..763C}.}
\label{tab:popIIIII}	
\end{table}
	
\subsection{Escape fraction }
\label{ssec:fesc}
Ionizing radiation is strongly absorbed by \HI and dust in the ISM. As a result, only a fraction $f_{\rm esc}(M, z)$ of the photons with energy $> 1$ Ryd produced in a halo of mass $M$ at redshift $z$ manage to escape into the IGM and contribute to reionization. 

Theoretical estimates of $f_\mathrm{esc}$ are difficult, as such quantity depends on complex details of the ISM structure in galaxies. At low redshift ($z < 4$) $f_\mathrm{esc}$ is experimentally constrained to be $<0.1$
\citep{2001ApJ...546..665S, 2006ApJ...651..688S, 2006MNRAS.371L...1I, 2007ApJ...668...62S, 2009ApJ...692.1287I, 2011ApJ...736...41B, 2012ApJ...751...70V, 2012MNRAS.424L..54V}.
As we will see later, in this redshift range the exact value is not very important because the results only depend on $(1-f_\mathrm{esc}) \approx 1$, but during EoR different values of $f_\mathrm{esc}$ lead to considerably different reionization histories.

To further complicate the above situation, results from numerical simulations are often inconsistent \citep{Dove00, Wood00, Gnedin08, Razoumov06, Wise09}, even if they generally suggest an higher $f_{\rm esc}$ for smaller galaxies. \cite{2013MNRAS.431.2826F} suggest a bimodal function, with $f_{\rm esc} \approx 1$ for the smallest galaxies and $f_{\rm esc} \approx 0$ for the others. In this work we will use this approach, tweaking the three free parameters (the two escape fractions for small and big galaxies and the threshold mass) to have a reasonable reionization history:
\begin{equation}
  f_{\rm esc}(M, z) = \left\{ \begin{array}{cc} 0.05 & {\rm for \ } M > 2\times 10^9 M_\odot \\  0.35 & {\rm for \ } M < 2 \times 10^9 M_\odot \\ \end{array} \right. . 
\end{equation}
Therefore the {\it effective} ionizing rate measured in the IGM from a given halo is $f_{\rm esc}(M, z)\dot N_{912}(M, z)$; the density rate of ionizing photons is then obtained from
\begin{gather}
\label{ndotion}
\dot n_\mathrm{ion} = \int_{M_{\rm min}} dM f_{\rm esc}(M, z) \dot N_{912}(M, z) \frac{dn}{dM};
\end{gather}
The mass-weighted $\langle {f_\mathrm{esc}}(z) \rangle_M$ approaches 0.35 at very high redshift because the mean mass of the halos decreases below the threshold mass. As we are going to see next, this behavior accelerates reionization and suppresses \Lya emission from low-mass galaxies.

\section{Reionization}
\label{ssec:reion}
A detailed treatment of reionization is not essential in our analysis, because, as we will see in \S \ref{ssec:meanigm}, recombinations from the ionized IGM are not the dominant source of Ly$\alpha$ photons. It could appear that the computation of the neutral fraction, $x_{\rm HI}$,  in the IGM is essential, because \Lya photons interact strongly with \HI. However, although numerical simulations indicate that \HI can reduce significantly the observed luminosity of individual galaxies \citep{2011MNRAS.410..830D, 2008MNRAS.386.1990M, 2007MNRAS.381...75M, 2014PASA...31...40D}, this effect is not relevant for \Lya intensity mapping. This is because Ly$\alpha$ photons are not destroyed by \HI, but are simply scattered; the net effect is a damping of fluctuations on scales below the typical photon mean free path, $\lambda \approx 1$ Mpc.

The diffuse IGM emission depends on the opacity of resonant Lyman photons which is derived from observations at $z < 5.5$ and extrapolated at higher redshifts. We use a simplified approach to compute the ionization state of the IGM that assumes a bimodal (two-phase) distribution in which the gas is either fully ionized or neutral; in other words, we ignore for the moment the residual \HI in the ionized regions. We will further discuss this issue in Sec. \ref{ssec:meancont}.

\begin{figure}
\vspace{+0\baselineskip} {\includegraphics[width=0.45\textwidth]{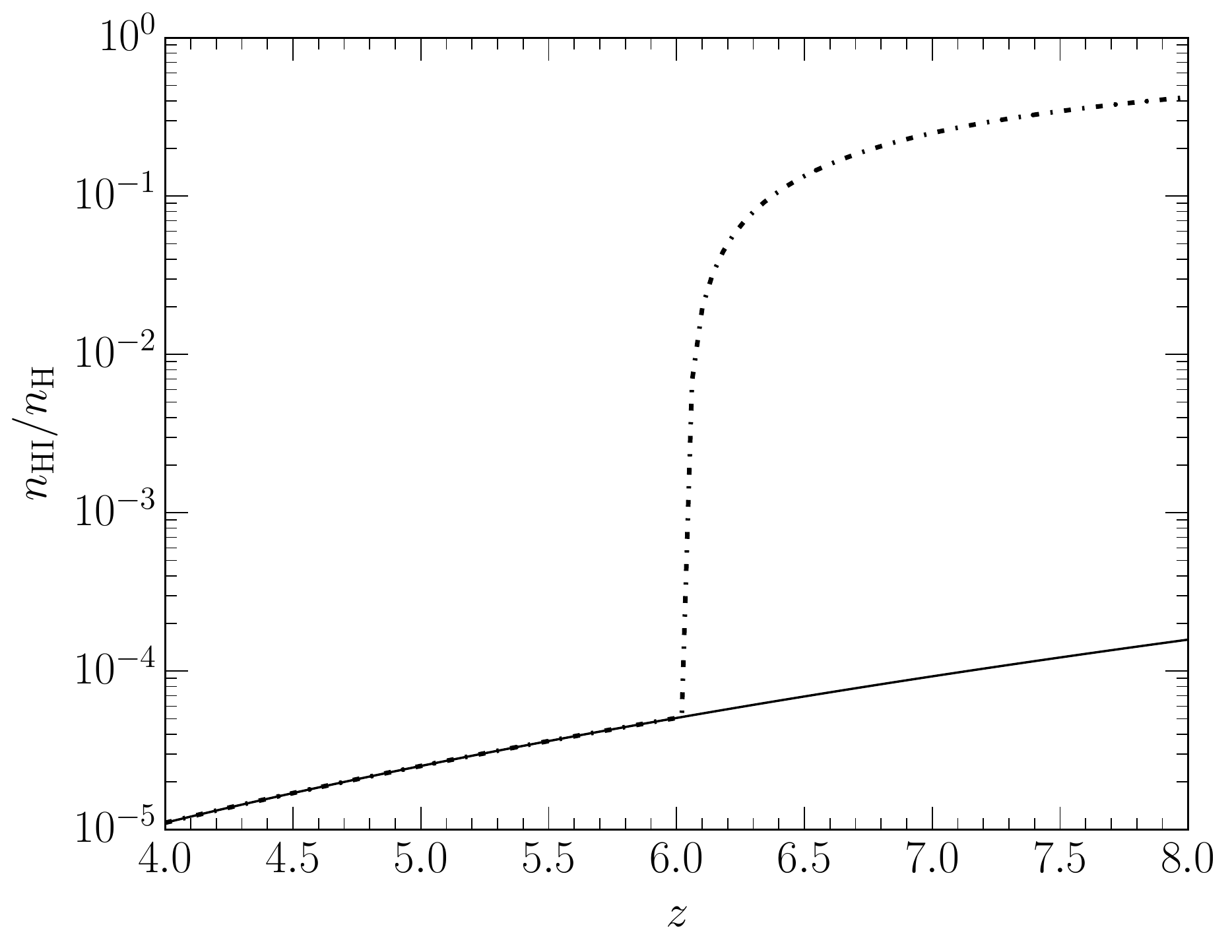}}
\caption{Redshift evolution of the neutral hydrogen fraction $x_{\rm HI} = n_{\rm HI}/n_H$ (dot-dashed line) and residual $x_{\rm HI}$ in the ionized regions (solid line). During reionization $x_{\rm HI} \simeq 1 - Q$; after its completion,  $x_{\rm HI}$ is computed from the ionization equilibrium condition matched to quasar absorption line data (see eq. \ref{efftau}).}
\vspace{-1\baselineskip}
\label{fig:fil}
\end{figure}

The purpose of this Section is to compute the filling factor $Q$, e.g. the volume fraction filled by ionized IGM. The governing evolutionary equation can be obtained by imposing a detailed balance condition among the number density of ionizing photons, ionizations and recombinations \citep{2013fgu..book.....L}:
\begin{equation}
\label{eqreion}
n_H \frac{dQ}{dt} = \dot n_{\rm ion} - n_H^2 (1+z)^3 C(z) \alpha_B(T) Q,
\end{equation}
where $C(z)$ is the fiducial clumping factor computed by \cite{2009MNRAS.394.1812P} at $z > 6$ (it was used the fit with $C_{100}$ and $z_r = 9.0$ from appendix A, with $C(z < 6) = C(6)$) and $\alpha_B(T)$ is the Case B recombination coefficient \citep{2009RvMP...81.1405M, 2013fgu..book.....L}. Fig. \ref{fig:fil} shows the numerical solution of eq. \eqref{eqreion}. We used the reionization history model to constrain the escape fraction; we obtained $Q = 1$ at $z \approx 6.0$, $Q = 0.9$ at $z\approx6.4$ and $Q = 0.5$ at $z=8.6$. The corresponding electron scattering optical depth is $\tau_{\rm es} = 0.067$, compatible with the most recent Planck result \citep{2015arXiv150201589P} of $0.066 \pm 0.016$. 

\section{Mean \Lya intensity}
\label{sec:meanint}
In this Section we will compute the mean intensity of the \Lya signal. The intensity, $I$, observed at frequency $\nu_{0}$ and redshift $z=z_0$ is simply the integral of the emissivity, $\epsilon$, along the cosmological path $\ell$ \citep{2013fgu..book.....L}:
\begin{equation}
\label{ip}
I(\nu_{0}, z_0) = \frac{1}{4 \pi} (1+z_0)^3 \int_{z_0}^{\infty} dz' \frac{d\ell}{dz} \epsilon\left[ \nu_{0} (1+z'), z' \right] ,
\end{equation}
where $d\ell/dz = c \left[(1+z)H(z)\right]^{-1}$. In this context the exact shape of Ly$\alpha$ line profile is not relevant as in practice the typical line width is negligible compared to the spectroscopic resolution of the instruments. For this reason we assume a Dirac delta-function, $\delta_D$. Hence 
\begin{equation}
\epsilon(\nu, z) = \dot \rho^\lya(z)\delta_D(\nu - \nu_{\alpha}),
\end{equation}
$\dot \rho^\lya$ is the \Lya luminosity density. In this case the integral reduces to
\begin{equation}
\label{meanI}
I(\nu_{0})= \frac{c}{4 \pi} \frac{\dot \rho^\lya(z_{\alpha})}{\nu_{\alpha} H(z_{\alpha})},
\end{equation}
where $z_{\alpha} = \nu_{\alpha}/\nu_{0} -1$ is the redshift of emission of the \Lya photons. We will consider three main processes leading to \Lya emission: recombinations in the ISM of halos/galaxies ($\dot \rho^\lya_{\rm halo}$), recombinations in the IGM ($\dot \rho^\lya_{\rm IGM}$), and excitations due to UV photons ($\dot \rho^\lya_{\rm cont}$).

\subsection{Recombinations in the ISM}
\label{ssec:meanrec}

When an electron recombines with a proton, it can directly populate the ground state (Case A) or an excited state (Case B); 
in the second case there is a chance of Ly$\alpha$ emission. Basically, an excited hydrogen atom can decay to the ground state in four ways (e.g. \cite{2014PASA...31...40D}): (i) from the $2S^2$ state, it can decay only through two-photon emission; (ii) from the $2P^2$ state a Ly$\alpha$ photon is emitted; (iii) decay from higher levels produces the emission of a Ly$n$ photon, or (iv) a cascade to another excited state and then one of the first three cases. Lyman photons interact strongly with hydrogen and we can assume the Ly$n$ photons to be locally absorbed and reemitted as long as they end up into a Ly$\alpha$ photon or in a two-photon emission. If we include all the four emission mechanisms, a fraction $f_{\alpha}$ of the Case B recombination emits a Ly$\alpha$ photon \citep{2008ApJ...672...48C}
\begin{equation}
f_{\alpha} =    0.686 - 0.106 \log T_4 - 0.009 T_4^{-0.44}, 
\end{equation}
where $T_4 = T/(10^4K)$.	As the temperature dependence is weak, we will assume in the following $T_4 = 2$ and hence $f_{Ly\alpha} = 0.65$.
 
As we have seen, a fraction $f_{\rm esc}(M, z)$ of the ionizing photons escape into the IGM, while the remaining part are absorbed by \HI photoionization. The recombination timescale for typical ISM densities ($n_H \approx 1 \, \mathrm{cm}^{-3}$) is $t_{rec} \approx 1/n_H\alpha_B \approx 2 \times 10^5 {\rm yr}$ and is negligible compared to the Hubble time. Thus we can assume photoionization equilibrium and that any \HI photoionization generates $f_\alpha$ \Lya photons.
	
However, the effects of dust absorption can be very significant especially at low redshift. Depending on the clumpiness of the ISM, the resonant nature of the \Lya line can lead to both a suppression \citep{2011ApJ...730....8H} or an amplification \citep{1991ApJ...370L..85N, 2006A&A...460..397V} of \Lya radiation with respect to the UV continuum.
To overcome these difficulties we resort to an empirical approach. \cite{2009A&A...506L...1A} found a relation between the \Lya escape fraction ($f^{\alpha}_{\rm esc}$) and the reddening $E(B-V)$ in low redshift LAE ($z \approx 0.3$).
\begin{equation}
\label{fescalpha}
f^{\alpha}_{\rm esc} = 10^{-0.4 k_{Ly\alpha} E(B-V) },
\end{equation}
where $k_\lya = 12.7$. We computed the $E(B-V)$ of each halo from the observational $\beta$ (eq. \eqref{attenuation}) and using the relation \citep{2000ApJ...533..682C}
\begin{equation}
\label{ebv}
 E(B-V)(M, z) = 0.44 A_{1600}(M, z)/ 4.39 . 
\end{equation}
\begin{figure}
\vspace{+0\baselineskip}
{
\includegraphics[width=0.45\textwidth]{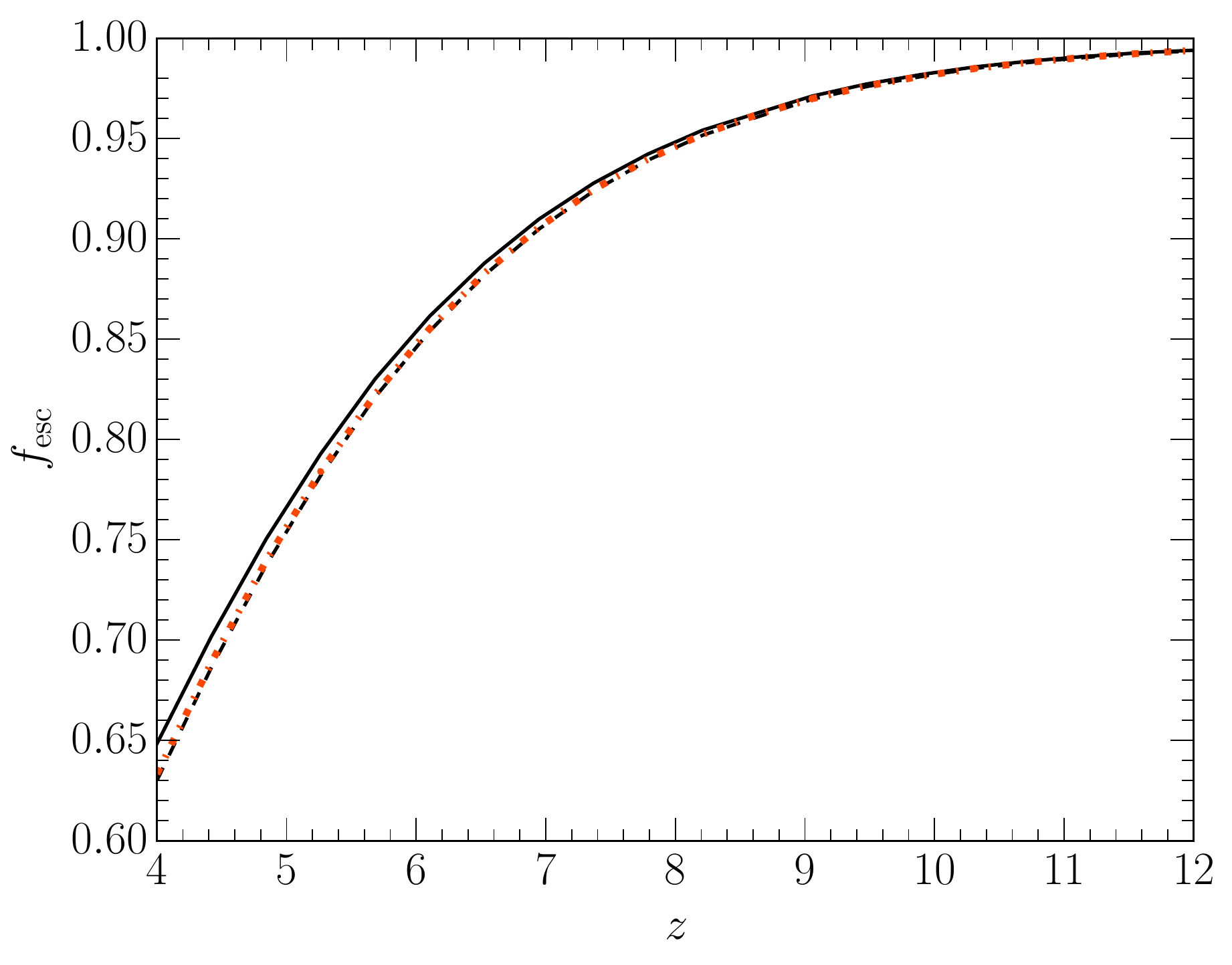}
}
\caption{Average \Lya escape fraction, $\langle f^\alpha_{\rm esc} \rangle_\lya$ from \protect\cite{2009A&A...506L...1A} (orange dash-dotted, eq. \eqref{fescalpha}) weighted with the \Lya luminosity (eq. \eqref{lyalum}); for comparison, the average Calzetti dust attenuation law \protect\citep{2000ApJ...533..682C} (eq. \eqref{calext}) at $912$\,\AA (black dashed) and $1216$\,\AA (black solid).}
\vspace{-1\baselineskip}
\label{fig:fdust}
\end{figure}
As a caveat we point out that the we are implicitly making two important assumptions:  (i) $f^{\alpha}_{\rm esc}$ depends only on the dust content, (ii) we can extrapolate results obtained for low-redshift systems to high-$z$. In spite of these uncertainties, the present approach has the advantage that it yields the extinction of \Lya photons by dust as a function of mass and consistently with the UV dust absorption. Fig. 4 shows the average \Lya escape fraction, $\langle f^\alpha_{\rm esc} \rangle_\lya$.   
  
Under these hypothesis, the halo luminosity is given by
\begin{multline}
\label{lyalum}
L_{Ly\alpha}(M, z) = h_P \nu_{Ly\alpha} f_{\alpha} \times\\
f^{\alpha}_{\rm esc}(M, z) (1 - f_{\rm esc}(M, z))\dot N_{912}(M, z); 
\end{multline}
the luminosity density is
\begin{equation}
\dot \rho^\lya_{\rm halo} = \int_{M_{\rm min}} d{M} \frac{d{n}}{d{M}} L_{\lya}(M, z), 
\end{equation}
And, finally, the intensity is
\begin{multline}
\label{Ihalo}
I^\lya_{halo}(z) =  \frac{c h_P f_{\alpha}}{4 \pi H(z)} \int_{M_{\rm min}} d{M} \frac{d{n}}{d{M}} f^{\alpha}_{\rm esc}(1 - f_{\rm esc})\dot N_{912}.
\end{multline}
Clearly, considerable uncertainties remains on the parameters in this formula: we have already discussed those concerning $f^{\alpha}_{\mathrm{esc}}$ and $f_{\rm esc}$; we also recall here that $\dot N_{912}$ is extrapolated to masses not probed by HST surveys. These uncertainties are amplified in the estimate of the PS,  which scales as ${I_\lya}^2$. Therefore, from this estimate we cannot safely predict the \Lya PS that should be observed, but we can conclude that a survey can constrain the mass-averaged values of ${f^{\alpha}_{\mathrm{esc}}}$, ${ f_{\rm esc}}$ and ${\dot N_{912}}$. Of course, these parameters are strongly degenerate.

\subsubsection{Lya luminosity function}
\begin{figure}
\vspace{+0\baselineskip}{\includegraphics[width=0.45\textwidth]{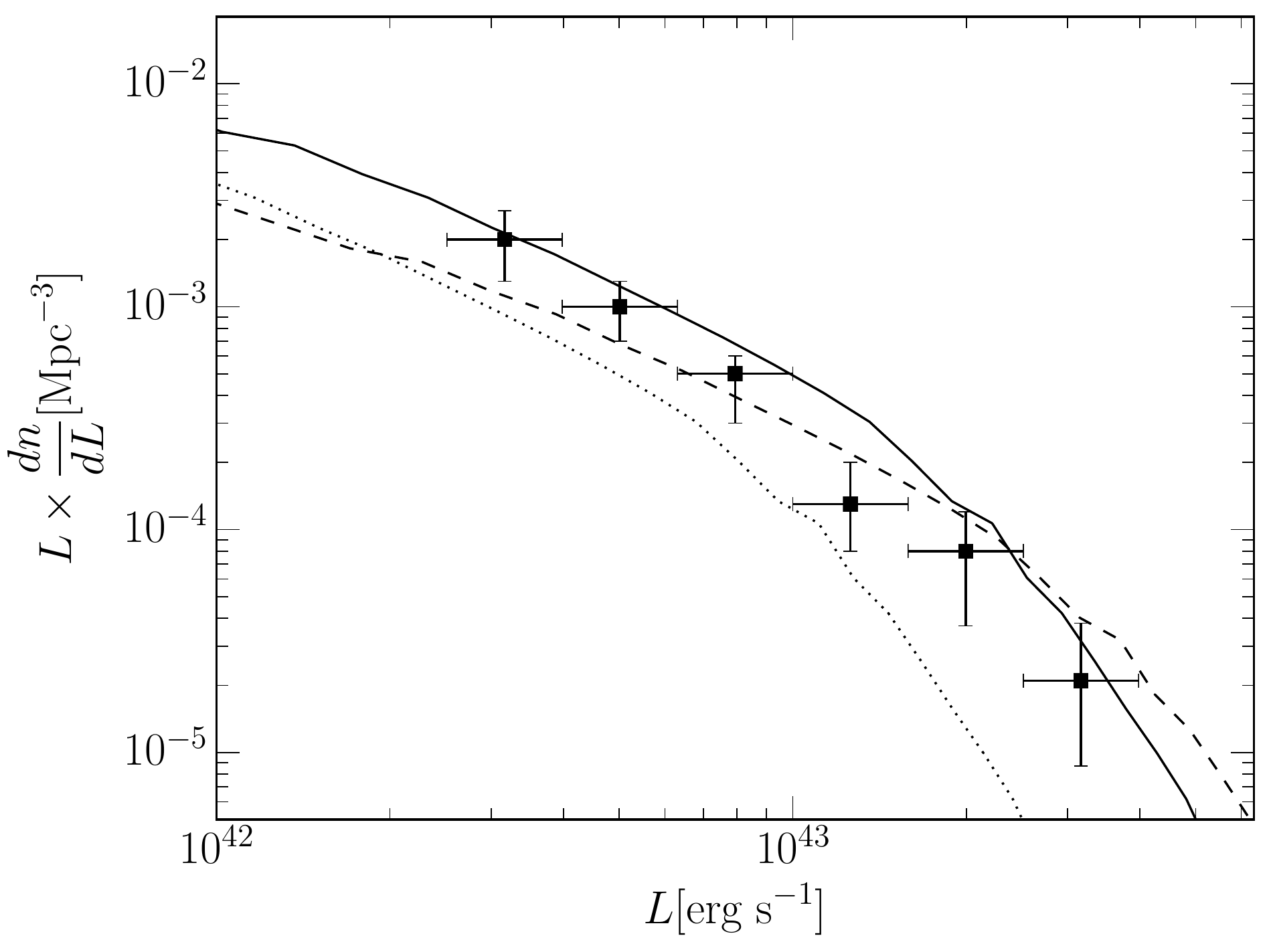}}
\vspace{-1\baselineskip}
\caption{\Lya luminosity function at $z=5.7$. We show our model (i) without IGM absorption or duty cycle (solid line); (ii) with a 50\% IGM absorption (dotted line); (iii) with also a 30\% duty cycle (dashed line). The experimental data are from \protect\cite{2008ApJS..176..301O}. The faintest galaxies in our model have luminosity $\approx 9\times 10^{38}$erg/s and therefore are not shown in this plot.}
\label{fig:LF}
\end{figure}

As a by-product of the method and a sanity check, we can predict the \Lya luminosity function (LF) of our galaxies. This is shown in    
Fig. \ref{fig:LF} for $z \approx 5.7$ (solid line), where it is also compared to the experimental data from \cite{2008ApJS..176..301O}; although the agreement seems adequate, we recall that we have not included the \Lya line damping \citep{2011MNRAS.410..830D,  2014PASA...31...40D} of the blue part of the line due to the intervening IGM. To provide a rough estimate of this effect we also show the LF in which a constant transmissivity factor $T_\alpha = 0.5$ (dotted line). Additionally, we note that if LAE are characterized by a duty cycle, this could also modify the shape of the LF at fixed $\dot \rho^\lya_{\rm halo}$. For example the dashed line in Fig. \ref{fig:LF} corresponds to a LF with: (1) $T_\alpha = 0.5$, (2) a 30\% duty cycle (i.e. only 30\% of the galaxies host a LAE), and (3) an increased (by a factor 3.3)  individual LAE luminosity.  Using more sophisticated (but still arbitrary) prescriptions for the duty cycle it should be possible to obtain a LF in perfect agreement with the observational data. For the present goals, we consider the accuracy of the present results as satisfactory.  %

\subsection{Recombinations in the IGM}
\label{ssec:meanigm}
In our reionization model, we approximate the IGM in a two-phase medium defined by its ionization stata (neutral/ionized).  Recombinations occur only in the ionized regions, and produce a luminosity density:
\begin{equation}
\dot \rho^\lya_{\rm IGM} =  h_p \nu_{\alpha} f_{\alpha} Q(z) n_H^2(0) (1+z)^3 C(z) \alpha_B(T);
\end{equation}
note the different dependence from the ionized fraction compared to the emission from a uniformly ionized gas which is $\propto Q^2(z)$ \citep{2014ApJ...786..111P}.

It follows that the intensity is 
\begin{equation}
\label{Irec}
I^\lya_{IGM}(z) =  \frac{c h_P}{4 \pi H(z)} f_{\alpha} Q(z) n_H(0) n_H(z) C(z) \alpha_B(T).
\end{equation}

\subsection{UV excitations in the IGM}
\label{ssec:meancont}

Continuum  UV photons with energy $10.2-13.6$ eV (i.e. between the \Lya line and the Lyman-limit) after escaping from galaxies redshift into Lyman-series frequencies and can be absorbed by \HI. The resulting excited \HI atoms eventually decay to the ground state, producing Ly$\alpha$ photons. The computation of the \Lya emissivity in this case is slightly more involved than for recombinations, as it requires a model for the production and absorption of $10.2-13.6$ eV photons \citep{2014ApJ...786..111P}.

The absorption probability of Ly$n$ photons at redshift $z$, $P_\mathrm{abs}(n, z)$, can be expressed by the canonical Gunn-Peterson \citep{1965ApJ...142.1633G} optical depth: 
\begin{equation}
\tau_{GP}(n, z) = \frac{3\gamma_n}{2H(z)} n_\mathrm{HI}(z) \lambda_{Lyn}^3,
\end{equation}
where 
\begin{equation}
\gamma_n = 50 \left(\frac{1-n^{-2}}{0.75}\right)^2\left(\frac{f_n}{0.4162}\right) \mathrm{MHz} 
\end{equation}
is the resonance half width at half-maximum, $f_n$ is the oscillator strength tabulated in \cite{1994A&AS..108..287V}, and $\lambda_{Lyn}$ is the wavelength corresponding to the $n$-transition.

A precise computation of $\tau_{GP}$ requires the knowledge of $n_{\rm HI}$ even in ionized regions. Our reionization model is too simple to compute it; but quasar absorption line observations provide the ``effective'' optical depth (i.e. the average of $\tau_{GP}$ over small scale inhomogeneities of overdensity $\Delta = \rho/\overline{\rho}$) for \Lya photons up to $z = 5.5$ \citep{2006AJ....132..117F}
\begin{equation}
\label{tauobs}
\tau^{\rm obs}_{\mathrm{eff}, \alpha} \equiv \langle \tau_{GP} (2,z) \rangle_\Delta = (0.85 \pm 0.06)\left( \frac{1+z}{5} \right)^{4.3\pm 0.3}.
\end{equation}
For $z > 5.5$ even the ionized IGM is optically thick to the \Lya line, which is the most relevant excitation (in the spectra from  \texttt{starburst99} the bulk of redshifted UV photons has a frequency between the Ly$\beta$ and the \Lya one). Our strategy is then to use a model for small scale inhomogeneities from which  we compute $\tau_{\mathrm{eff}, \alpha}$ as a function of $n_\mathrm{HI}$ in ionized regions. Then $n_{\rm HI}$ is tuned to reproduce $\tau^{\rm obs}_{\mathrm{eff}, \alpha}$.	The extrapolation to higher redshifts is arbitrary but not relevant for the results of this work, because the absorption is already saturated at $z = 5.5$ in the \Lya line.

To understand how $\tau_{GP}$ depends on IGM (over)density $\Delta$  we impose photoionization equilibrium \citep{2005MNRAS.361..577C}, an excellent assumption under IGM conditions: 
\begin{equation}
\label{ioneq}
n_\mathrm{HI}\Gamma = n_\mathrm{HII}n_e \alpha_B(T) C(z) ,
\end{equation}
with $n_\mathrm{HI} = (1-x_e)n_H$, $n_\mathrm{HII} \approx n_e = x_e n_H$, $\alpha_B(T) \propto T^{-0.7}$. 

As we are concerned with ionized regions in the post-reionization epoch, $x_e \approx 1$; in addition, we use an equation of state $T \propto \Delta^{\gamma -1}$, where $1 < \gamma < 2.4$ is the adiabatic index. The results depend weakly on $\gamma$ and we assume an equation of state with $\gamma = 1.4$. With these assumptions, $\tau_{GP} \propto \Delta^{2.7 - 0.7\gamma}$, i.e. $\tau_{GP} \propto \Delta^{1.72}$ for $\gamma=1.4$. 

As a model for the small scale inhomogeneities of the IGM, we adopt the one proposed by \cite{2000ApJ...530....1M}. Such model, tested against numerical simulations, gives the Probability Distribution Function for the overdensity $\Delta$ as
\begin{equation}
\label{miralda}
P(\Delta, z) = A(z) \mathrm{exp}\left( -\frac{[\Delta^{2/3}-C_0(z)]^2}{2[2\delta_0(z)/3]^2} \right) \Delta^{-\beta},
\end{equation}
where the constants $A, \beta, C_0$ are reported in Table \ref{tab:miralda} as a function of redshift and $\delta_0(z) = 7.61/(1+z)$. 
\begin{table}
\centering
\begin{tabular}{@{}lccr@{}}
\toprule
 $z$ & $A$  & $\beta$ & $C_0$ \\
\midrule
2 & 0.170 & 2.23 & 0.623 \\
3 & 0.244 & 2.35 & 0.583 \\
4 & 0.322 & 2.48 & 0.578 \\
6 & 0.380 & 2.50 & 0.868 \\
8 & 0.463 & 2.50 & 0.964 \\
10 & 0.560 & 2.50 & 0.993 \\
12 & 0.664 & 2.50 & 1.001 \\
15 & 0.823 & 2.50 & 1.002 \\
\bottomrule
\end{tabular}
\caption{Parameters for the IGM overdensity probability distribution function, $P(\Delta, z)$, used in eq. \protect\eqref{miralda}; data are from \protect\cite{2000ApJ...530....1M}}
\label{tab:miralda}	
\end{table}

The effective optical depth can now be computed as
\begin{equation}
 e^{-\tau_\mathrm{eff}(n, z)} = \langle e^{-\tau} \rangle_\Delta = \int_0^{\Delta_h} P(\Delta, z) e^{-\Delta^{1.72} \tau_{GP}(n, z)} d\Delta.
\label{efftau}	
\end{equation}
We have set $\Delta_h=100$ as a reasonable overdensity of the external parts of virialized halos; the results are however insensitive to the precise choice of this parameter. 
Using the least square method and eq. \eqref{tauobs} between $z=4$ and $5.5$ we find $x_{\rm HI} = 1.1 \times 10^{-5} [(1+z)/5]^{4.53}$ (Fig. \ref{fig:fil}) and extrapolate this relation to $z > 5.5$.  Given that the bulk of the continuum photons is emitted between the \Lya and the Ly$\beta$ line and that the absorption probability of \Lya photons, $P_{\rm abs}^\lya$, saturates at $z = 5.5$, the final results do depend only weakly on this extrapolation. 

If we consider the neutral IGM ($x_e \ll 1$), we have instead $\tau_{GP} \propto n_{\rm HI} \propto \Delta$, or
\begin{equation}
\label{taugpneut}
 e^{-\tau_\mathrm{eff, neut}(n, z)} = \int_0^{\Delta_h} P(\Delta, z) e^{-\Delta \tau_{GP}(n, z)} d\Delta;
\end{equation}
this results in $P_{\rm abs} = 1$ for the lowest Lyman levels.

Finally, the transmissivity of a Ly$n$ photon through the IGM at redshift $z$, $T(n, z) = 1 - P_{\rm abs}(n, z)$, is:
\begin{equation}
\label{trasm}
T(n, z) = Q e^{-\tau_\mathrm{eff}(n, z)} + (1-Q) e^{-\tau_\mathrm{eff, neut}(n, z)}.	
\end{equation}
\begin{figure}
\vspace{+0\baselineskip}
{\includegraphics[width=0.45\textwidth]{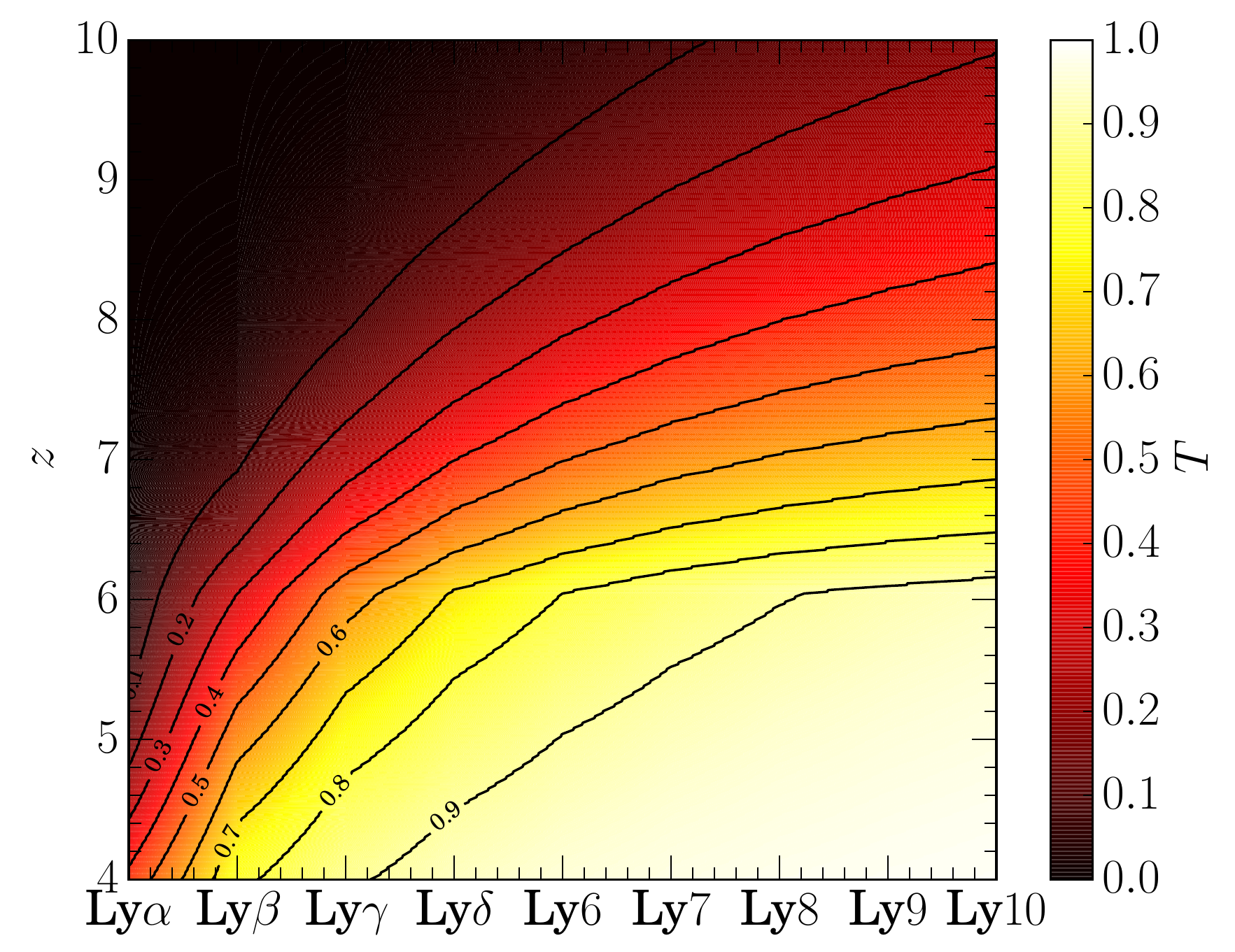}		}
\caption{Transmissivity of different Lyman-series lines through the IGM as a function of redshift. During the EoR the \Lya line saturates and is completely absorbed.}	
\vspace{-1\baselineskip}
\label{fig:pabs}
\end{figure}	
Fig. \ref{fig:pabs} shows the above result in graphical form for different Lyman-series lines. It is evident that during EoR the IGM is optically thick at least to \Lya photons, therefore all continuum photons are scattered and re-emitted as \Lya photons. On the other hand, the increasing transparency of the IGM at $z \simlt 6-7$, where $T(n, z)$ approaches unity, has important consequences that were not properly considered in earlier works \citep{2014ApJ...786..111P}.

A fraction $f(n)$ of the excited \HI atoms produce Ly$\alpha$ photons (see Table 1 from \cite{2006MNRAS.367..259H}), therefore the Ly$\alpha$ intensity at redshift $z$ is
\begin{equation}
I(\nu_\lya, z) = \sum_{n=2}^{\infty}\frac{h_P \nu_{Ly\alpha}}{h_P \nu_{Lyn}} P_\mathrm{abs}(n,z) f(n) I(\nu_{Lyn}, z) ;
\end{equation}
Then, from the radiative transfer equation \citep{2013fgu..book.....L} the intensity observed at $z=0$ and $\nu = \nu_\lya/(1+z)$ is
\begin{equation}
\label{Icontprec}
I^\lya_{\mathrm{cont}}(z) = I(\nu_\lya, z)/(1+z)^3 .
\end{equation}

Hence, the final step is to compute $I(\nu_{Lyn}, z)$; this is easily done once the UV emissivity, $\epsilon(\nu)$, is assigned.
We will parametrize the specific emissivity as:
\begin{multline}
\epsilon(\nu, z) = h_P \nu \dot n_\nu(\nu, z) =\\
= \int_{M_{\rm min}} d{M} \frac{d{n}}{d{M}} L_\nu(M, z) f^{\rm UV}_{\rm dust}(M, z) ,
\end{multline}	
where $L_\nu(M, z)$, the spectral luminosity density, is computed with \texttt{starburst99} coherently with Sec. \ref{ssec:sfrd}, and $f^{\rm UV}_{\rm dust}$ is the fraction of UV photons that escape unimpeded by dust. The latter can be accounted by using, e.g. the Calzetti extinction law \citep{2000ApJ...533..682C, 2008ApJ...685.1046L} in parametric form:
\begin{equation}
\label{calext}
f^{UV}_{\rm dust}(M, z) = 10^{-0.4 k_\lambda E(B-V)(M, z)},
\end{equation}
where $E(B-V)(z)$ is given in eq. \eqref{ebv}.
The redshift evolution of $\langle f^{UV}_{\rm dust}(z) \rangle_\lya$ is plotted in Fig. \ref{fig:fdust} for non-resonant $1216\, \text{\AA}$ photons and for $912\,\text{\AA}$ ones.
In this paper there are not significant differences between the dust attenuation of resonant \Lya photons and the neighbouring ones. 

For the calculation of the intensity, in this case we have also to consider absorption from \HI; 
indeed a photon, while redshifting into a Ly$n$ frequency, can pass through other Ly$m$ frequencies, with $m > n$. 
Therefore we will include a product of the transmission fractions from eq. \eqref{trasm}:
\begin{multline}
I(\nu_{Lyn}, z) = h_P \nu_{Lyn} \frac{(1+z)^2}{4 \pi} \int_z^{z_{max}(n, z)} dz' \times \\ 
\times \frac{c \dot n_\nu(\nu', z')}{H(z')} \prod_{n' = n+1}^{n'_{max}(n, z, z')} T(n', z_{n'}(\nu', z')),
\end{multline}		
where (i) $\nu' = \nu_{Lyn} (1+z')/(1+z)$ is the frequency of a photon at $z'$ that redshifts to $\nu_{Lyn}$ at $z$;
(ii) $1 +z_{max}(n, z) = ({\nu_{LL}}/{\nu_{Lyn}})(1+z)$ is the redshift at which a Lyman limit photon redshifts into a Ly$n$ photon at $z$;
(iii) $n'_{max} = \left[ 1-({\nu_{Lyn}}/{\nu_{LL}}) ({1+z'})/({1+z})\right]^{-1/2}$ is the maximum $m$ such that $\nu' > \nu_{Lym}$, and (iv) $1 + z_{n'} = ({\nu_{Lyn'}}/{\nu'})(1+z')$ is the redshift at which a photon with frequency $\nu'$ from redshift $z'$ redshifts into the Ly$n'$ frequency.
Thus from eq. \eqref{Icontprec}:
\begin{multline}
\label{Icont}
I^\lya_{\mathrm{cont}}(z) = \frac{c h_P \nu_{Ly\alpha}}{4 \pi (1+z)} \sum_{n = 2}^{\infty} P_{abs}(n, z) f(n)  \times \\
\times \int dz' \frac{\dot n_\nu(\nu', z')} {H(z')} \prod_{n' = n+1}^{n'_{max}} T(n', z_{n'}).
\end{multline}
As in Sec. \ref{ssec:meanrec}, in eq. \eqref{Icont} there are several poorly constrained parameters, both in $\dot n_\nu$ and in the IGM modelling. Therefore, if UV excitation dominates the \Lya PS, its measurement can constrain the intrinsic emission of UV photons, dust obscuration and the ionization state of the IGM. As for the halo emission, these three physical processes are degenerate and cannot be disentangled. 

As a final remark, we note a subtle point concerning photons scattered directly by the \Lya line: if the IGM were perfectly transparent, we would observe exactly the same intensity at the redshifted \Lya frequency. The effect of the IGM is to substitute part of the resonant \Lya photons from the line of sight (LOS) with the local average over the solid angle:
\begin{equation}
I_{\nu_\alpha}(\Omega_{\rm LOS}) \rightarrow P_{\rm abs} \langle I_{\nu_\alpha} \rangle_\Omega + (1-P_{\rm abs}) I_{\nu_\alpha}(\Omega_{\rm LOS}).
\end{equation}
In a perfectly homogeneous universe these two terms cancel each other out; however, for spatial fluctuations the cancellation does not occur and also the continuum emission along the LOS is suppressed by the foreground removal. Fig. \ref{fig:intintIII} shows the difference in \Lya intensity when the IGM \Lya scattering is either included or excluded.

\subsection{Additional processes}
\label{ssec:meanother}
Atoms can emit \Lya photons also when they are collisionally excited and emit decaying to the ground state. This mechanism needs both thermal kinetic energies comparable to the \Lya energy and the presence of \HI. Such conditions can be reasonably found only in the transition zones between ionized and neutral regions. Thus, the emissivity depends strongly on the morphology and temperature of the ionized regions, which is hard to model analytically. Moreover, as the excitation temperature of the \Lya line, $T_\alpha = 1.2 \times 10^5$ K is much larger than those typical of photo-heated gas ($\sim 1-5\times 10^4$ K), the abundance of energetic electrons is exponentially suppressed.  \cite{2008ApJ...672...48C} computed the collisional excitation coefficient for \Lya emission $q^{\rm eff}_{Ly\alpha}$ (along with the equivalent one for recombinations $\alpha^{\rm eff}_{Ly\alpha} = f_{Ly\alpha} \alpha_B$)
\begin{equation}
\label{canta}
\dot n_{Ly\alpha} = n_e n_{\rm HII} \alpha^{\rm eff}_{Ly\alpha} + n_e n_\mathrm{HI} q^{\rm eff}_{Ly\alpha}.
\end{equation}
As expected, $ q^{\rm eff}_{Ly\alpha}$ depends strongly on temperature: at $T = 2\times 10^4$K it is three orders of magnitude larger than $\alpha^{\rm eff}_{Ly\alpha}$. Therefore at this temperature collisional excitations are relevant only if $x_{\rm HI} \simgt 10^{-3}$; however, a factor 2 change in temperature results in a variation of three orders of magnitude in $q^{\rm eff}_{Ly\alpha}$. \cite{2013ApJ...763..132S} assumed the collisional excitation intensity to be 10\% of that from recombinations; \cite{2014ApJ...786..111P} found it negligible and did not consider it in their model. To avoid the introduction of poorly constrained parameters related to the morphology and the temperature of the gas, we make the most conservative choice and neglect this contribution.

An additional ionization source could be represented by exotic annihilating/decaying dark matter particles, that partially ionize the IGM almost uniformly. In this case the additional \Lya intensity from the IGM is  
\begin{equation}
I^\lya_{\mathrm{uni}}(z) = \frac{1-Q(z)}{Q(z)} \chi(z)^2 \left( \frac{2\times 10^4 K}{T_{\mathrm{uni}}} \right)^{0.7} I_{\mathrm{IGM, rec}}(z).
\end{equation}	

The temperature dependence ($T_{\mathrm{uni}}$) comes from the recombination coefficient $\alpha_B$ and introduces a potentially large factor for a luke-warm gas heated by X-ray photons; $\chi(z)$ is the ionized fraction outside the reionized bubbles. Given the uncertainties in the nature of the dark matter particle and annihilation cross-sections, we will not consider this source in detail; however, we point out that it might be significant before the completion of reionization (Fig. \ref{fig:intintIII}).
	
\subsection{Total intensity}
\label{ssec:meantot}
\begin{figure}
\vspace{+0\baselineskip}
{\includegraphics[width=0.45\textwidth]{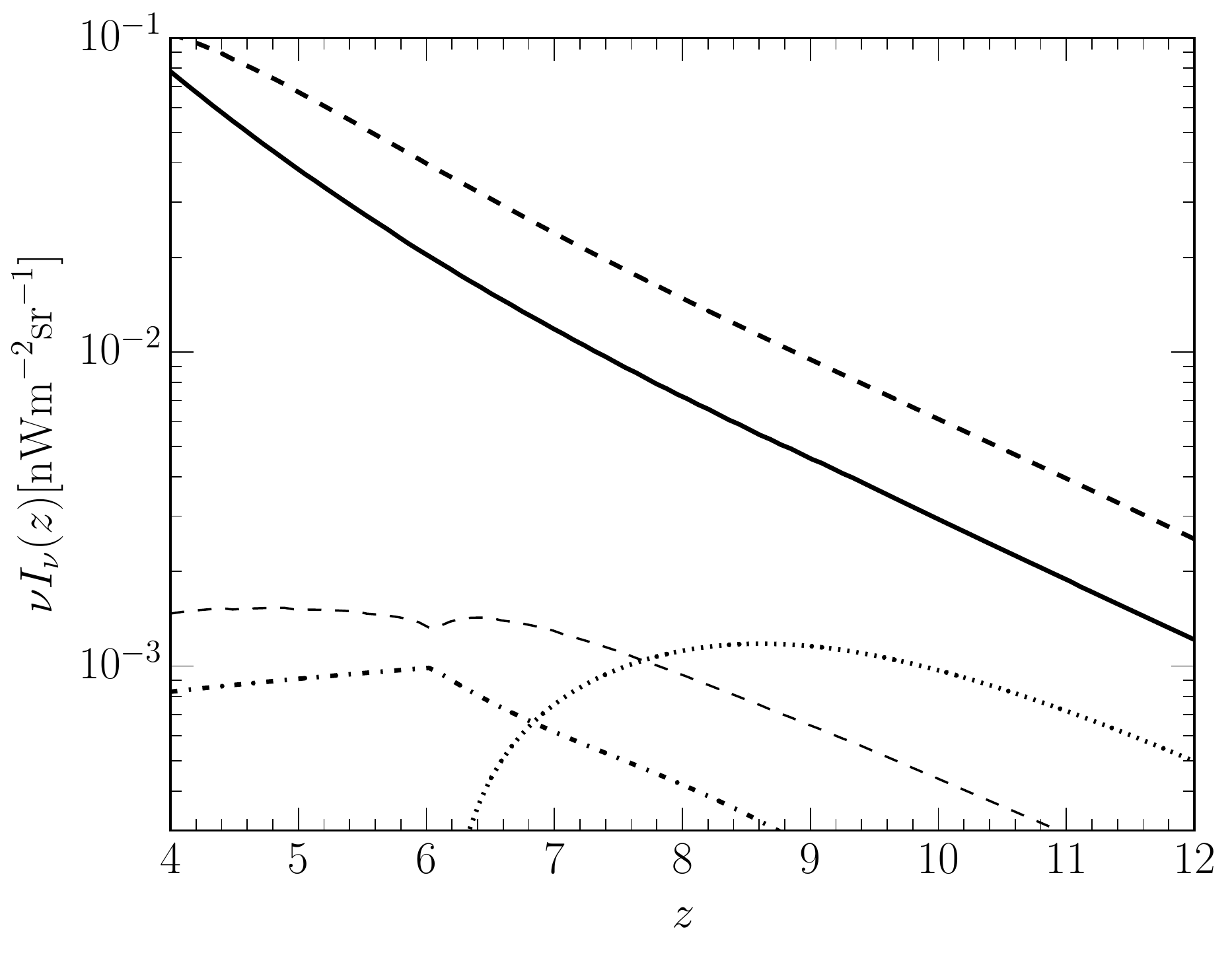}}
\vspace{-1\baselineskip}
\caption{\Lya mean intensities. Each curve represents the contribution from different physical processes:  (i) recombinations in the ISM (solid), (ii)  recombinations in the IGM (dot-dashed), (iii)  UV excitations in the IGM (thick dashed), (iv) same as (iii) but without photons scattered in the \Lya line (thin dashed), (v) emission from an hypothetical uniformly ionized  cold IGM ($x_{\rm HI} = 0.3$, $T_{\rm IGM} = 100$K, dotted).  }
\label{fig:intintIII}
\end{figure}

\begin{figure}		
\vspace{+0\baselineskip}
{
\includegraphics[width=0.45\textwidth]{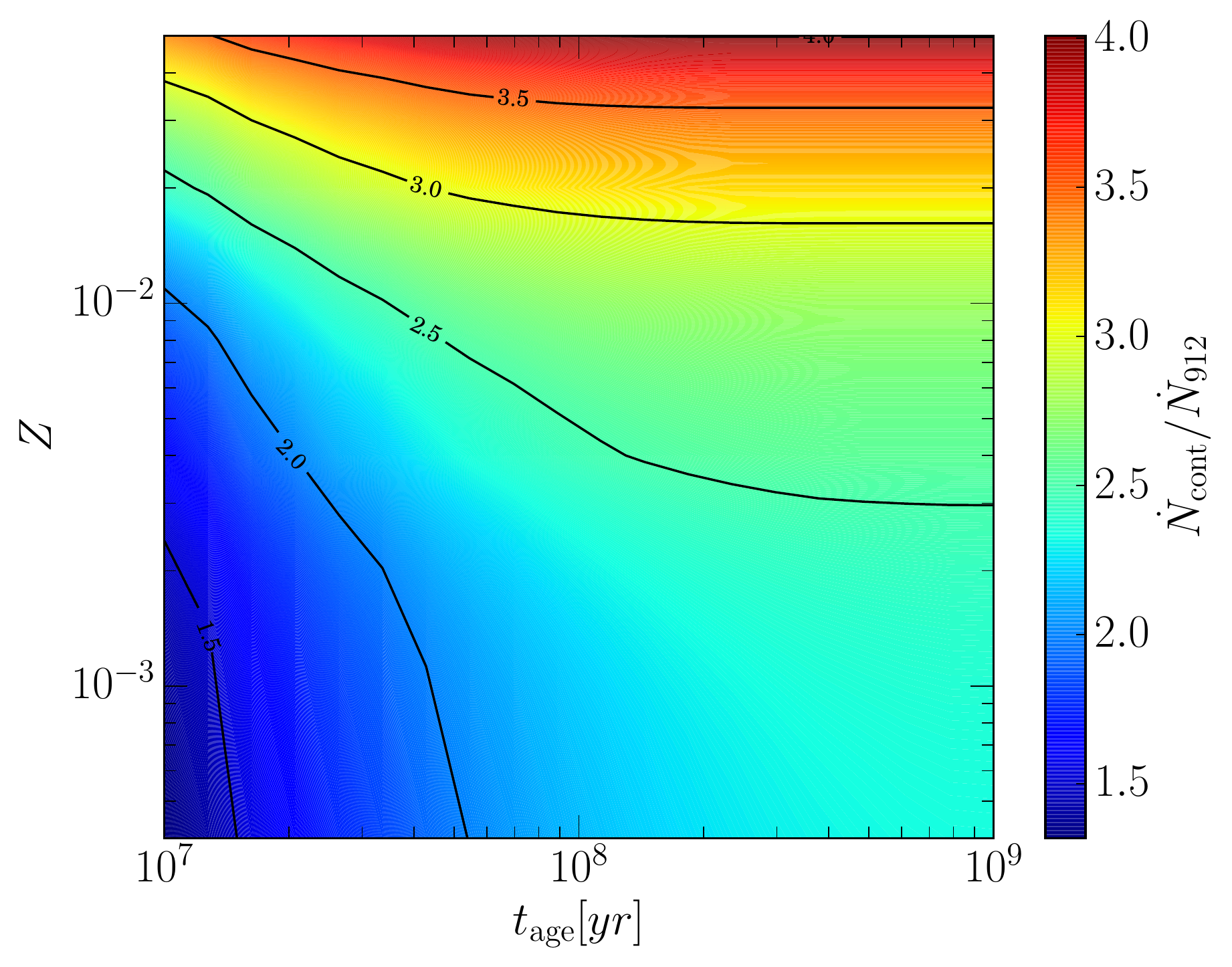}
}
\caption{Ratio between the emission rates of $E=10.2-13.6$ eV and ionizing photons. A constant SFR and a Salpeter IMF between 0.1 and 100 $M_\odot$ has been assumed.}
\vspace{-1\baselineskip}
\label{fig:phem}
\end{figure}

Fig. \ref{fig:intintIII} show the results of the previous estimates.
The dominant source is represented by the scattering of continuum photons. This can be understood from the fact that stars emit more UV photons with $10.2{\rm eV} < E < 13.6{\rm eV}$ than ionizing photons (see Fig. \ref{fig:phem}). The efficiency of these UV photons in producing Ly$\alpha$ photons is comparable to the ionizing ones and the effect of dust absorption is of the same order of magnitude. However, as we will see in the next Section, fluctuations induced by UV photon scattering have smaller amplitudes on small scales, because while UV photons originate from galaxies, Ly$\alpha$ photons are eventually produced in more distant regions of the IGM. 

Recently \cite{2015arXiv150404088C} made the first intensity mapping observation by cross-correlating the spectra from the SDSS with the QSO distribution at $z = 2.55$. They found a \Lya comoving luminosity density of $\dot \rho^\lya_{\rm SDSS} \approx 3.1 \times 10^{41} (3/b_\alpha) {\rm erg~}{\rm s}^{-1}{\rm Mpc}^{-1}$ ($b_\alpha = 3$ is their fiducial mean \Lya bias), with statistical uncertainties of $\sim 25\%$. Our model predicts $\dot \rho^\lya_{\rm IGM} = 8.73(6.51)\times 10^{40}{\rm erg~}{\rm s}^{-1}{\rm Mpc}^{-1}$ and $\dot \rho^\lya_{\rm ISM} = 6.62(3.21) \times 10^{40}{\rm erg~}{\rm s}^{-1}{\rm Mpc}^{-1}$ at $z = 4(7)$. As we will see in Sec. \ref{ssec:flucttot}, in this context only the ISM emission is important, since the IGM emission is smooth on the scales considered by \cite{2015arXiv150404088C} ($ < 100$ Mpc). If we linearly extrapolate our results to $z = 2.55$, we find $\dot \rho^\lya_{\rm ISM}(z = 2.55) \approx 0.91 \times 10^{41}{\rm erg~}{\rm s}^{-1}{\rm Mpc}^{-1}$, a value of the same order of magnitude of the \cite{2015arXiv150404088C} results, albeit a factor of 3 smaller.     

\section{Intensity Fluctuations}
\label{sec:fluct}
The observation of the mean intensity of \Lya emission is severely limited by foregrounds: the intensity of the NIR foreground around $1\,\mu$m is of about $10^{3-4} \,{\rm nW}{\rm \, m}^{-2}{\rm \, sr^{-1}}$ \citep{2005PhR...409..361K}, while our predicted \Lya signal is 5-6 orders of magnitude fainter. On the other side the situation is different if we consider the fluctuations, because the dominant foreground components are smooth in frequency and can be removed \citep{2006ApJ...650..529W, 2015arXiv150104429C, 2015arXiv150103823W}.

In this Section we will develop the theory necessary to predict the power spectrum of the fluctuations with our model. As for the intensity we will examine the three main \Lya photon emission processes (ISM emission, IGM recombinations and IGM Lyman absorption) separately.

\subsection{Fluctuations from ISM recombinations}
\label{ssec:fluctism}
The distribution of galaxies in space fluctuates due to: (i) clustering of large scale structure, and (ii) shot noise due to the discrete nature of the sources. 

The clustering term can be written in terms of the halo mass function. In the linear regime, $\delta(dn/dM) \approx b(dn/dM)\delta$ \citep{1999MNRAS.308..119S, 2001MNRAS.323....1S}. Thus,
\begin{equation}
\delta I^\lya_{\rm halo}(z, \mathbf{x}) = I^\lya_{\rm halo}(z) \langle b(z) \rangle_\alpha \delta(\mathbf{x}),
\end{equation}
where $\langle b(z)\rangle_\alpha$ is the mean bias weighted with the Ly$\alpha$ luminosity and the mass function (Fig. \ref{fig:biasfil}):
\begin{equation}
\label{meanbias}
\langle  b(z) \rangle_\alpha=\frac{\int_{M_\mathrm{min}}^{+\infty} L_{{\rm Ly}\alpha}(M, z) \frac{dn}{dM} b(M, z)}{\dot \rho^\lya_{\rm halo}}.
\end{equation}
For simplicity we have neglected redshift space distortions, because the interplay between peculiar velocities \citep{1987MNRAS.227....1K} and radiative transfer effects \citep{2011ApJ...726...38Z, 2015arXiv150404088C} is unclear. This is a conservative choice because this effect increases the linear bias \citep{2014ApJ...785...72G, 2015arXiv150103177V}. Therefore the Ly$\alpha$ power spectrum from galaxies is proportional to the linear power spectrum, $P_{gg} = \langle {b} \rangle_\alpha^2\sigma_k^2$  (we used the transfer function from \cite{1998ApJ...496..605E}). It follows that 
\begin{equation}
\label{pshlin}
P^\lya_\mathrm{h,\delta\delta} = I^\lya_\mathrm{halo}(z)^2 \langle b(z) \rangle_\alpha^2 \sigma_k^2(z).
\end{equation}
We have also neglected fluctuations associated with the distribution of satellite sources within a single halo (ofter referred to as the two-halo term) because on those scales ($k > h \mathrm{Mpc}^{-1}$) the power spectrum is completely dominated by shot noise.	
	
The second source of fluctuations, shot noise, is due to the Poisson fluctuations of the number of galaxies in a volume $V$ around the mean value,
\begin{equation}
\overline N_M = V \frac{dn}{dM}dM \qquad \mathrm{with}\,\, M \in [M, M+dM],
\end{equation}
and with standard deviation $\langle \delta N_M ^2\rangle =  \sigma_{N_M}^2 = \overline N_M$. To compute the power spectrum we can imagine to divide the mass range into a large number of bins of width $dM$ centered at $M_i$ ($i \gg 1$ is an integer index). As different bins are uncorrelated, and recalling that 
\begin{equation}
\delta_\mathrm{SN}(i, \mathbf{x}, dM) = \frac{\delta N_{M_i}}{V}, \\
\end{equation}
we can directly write the expression for the luminosity density fluctuations:
\begin{equation}
\delta \dot \rho^\lya_{\rm halo}(z, \mathbf{x}) = \dot \rho^\lya_{\rm halo}(z) \frac{\sum_i L_{{\rm Ly}\alpha}(M_i) \delta_\mathrm{SN}(i, \mathbf{x}, dM)}{\dot \rho^\lya_{\rm halo}(z)}.
\end{equation}
More often, though, we will be interested in the correlation function. Starting from
\begin{gather}
\langle \delta_\mathrm{SN}(i, \mathbf{x}, dM) \delta_\mathrm{SN}(j, \mathbf{x'}, dM) \rangle = \left( \frac{dn}{dM} \right) \delta_{ij}\delta_D^3(\mathbf{x} - \mathbf{x'})dM,
\end{gather}
we then find 
\begin{multline}
\langle \delta \dot \rho^\lya_{\rm halo}(z, \mathbf{x}) \delta {\dot \rho^\lya_{\rm halo}}(z, \mathbf{x'})\rangle =\\
 = {\dot \rho^\lya_{\rm halo}}(z)^2\frac{\sum_{i} L_{{\rm Ly}\alpha}(M_i)^2 \delta_D^3(\mathbf{x} - \mathbf{x'}) \frac{dn}{dM} dM }{{{\dot \rho}^\lya_{\rm halo}}(z)^2}.
\end{multline}
Finally we take the continuum limit $dM \rightarrow 0$ to obtain the shot-noise power spectrum. Since $\langle \delta \dot \rho^\lya_{\rm halo}(z, \mathbf{x}) \delta {\dot \rho^\lya_{\rm halo}}(z, \mathbf{x'})\rangle \propto \delta_D^3({\bf x} - {\bf x'})$, the power spectrum is independent of $k$, for the lack of spatial correlations:
\begin{equation}
\label{pshsn}
P^\lya_\mathrm{h,SN}(z) = I^\lya_\mathrm{halo}(z)^2 \frac{\int_{M_\mathrm{min}}^{M_\mathrm{max}} dM \frac{dn}{dM} L_{{\rm Ly}\alpha}(M, z)^2}{{\dot \rho^\lya_{\rm halo}(z)}^2};
\end{equation}
%

Shot noise dominates the signal for $k > 1 h{\rm Mpc}^{-1}$, but on these scales the luminosity from the point sources is smoothed by the scatterings in the CGM/IGM, suppressing the power spectrum. Thus, while measurements of the clustering component on scales $k < 1 h {\rm Mpc}^{-1}$ are unaffected by this problem, detections of the Ly$\alpha$ signal, as proposed by e.g. \cite{2014ApJ...786..111P}, might run into difficulties as the bulk of the predicted signal to noise originates from uncertain high-$k$ fluctuations. It is very difficult to investigate this effect analytically, because it depends both on the distribution of the relic \HI in the ionized regions and on the size of the ionized bubbles.  As we have not modeled these aspects in detail, our shot noise results should be used with caution.

\subsection{Fluctuations from IGM recombinations}
\label{ssec:fluctigm}
IGM emission is powered by hydrogen recombinations; to compute its fluctuations we then need to start from the hydrogen density field, $\delta_H$. Although assuming that baryons trace dark matter is a common practice, this hypothesis becomes increasingly inaccurate on small scales, where gas pressure can smooth fluctuations. This effect can be appropriately described by a simple model from \cite{2005ApJ...630..643M}, prescribing that at $z \simlt 8$ fluctuations are truncated for $k > k_F(z)$:
\begin{equation}
P_H(k, z) = F_b(k/k_F(z)) \sigma^2_k
\end{equation}
\begin{equation}
F_b(x) = \frac{1}{2}e^{-x^2} + \frac{1}{2(1+4x^2)^{1/4}},
\end{equation}
where $k_F(z) = 34 \Omega_m(z)^{1/2} h \mathrm{Mpc}^{-1}$ and $F_b(x) = 1$ for $z \ge 8$.
This correction, however, turns out to be irrelevant for our work because we will be mainly concentrated on large scale structure signales, i.e. fluctuations with $k \approx 0.1 h\mathrm{Mpc}^{-1}$, and therefore $\delta_H \approx \delta$.

The intensity from IGM recombinations is proportional both to the square of the hydrogen density and to the filling factor (eq. \eqref{Irec}):
\begin{equation}
I^\lya_\mathrm{IGM} \propto Q(z)n_H^2;
\end{equation} 
These two terms generate a different types of fluctuations: hydrogen density fluctuations are the only ones present after the completion of reionization;  those induced by the patchy reionization morphology dominate at high-redshift, as ionized regions are biased.

\begin{figure}
	\vspace{+0\baselineskip}
	{
	\includegraphics[width=0.45\textwidth]{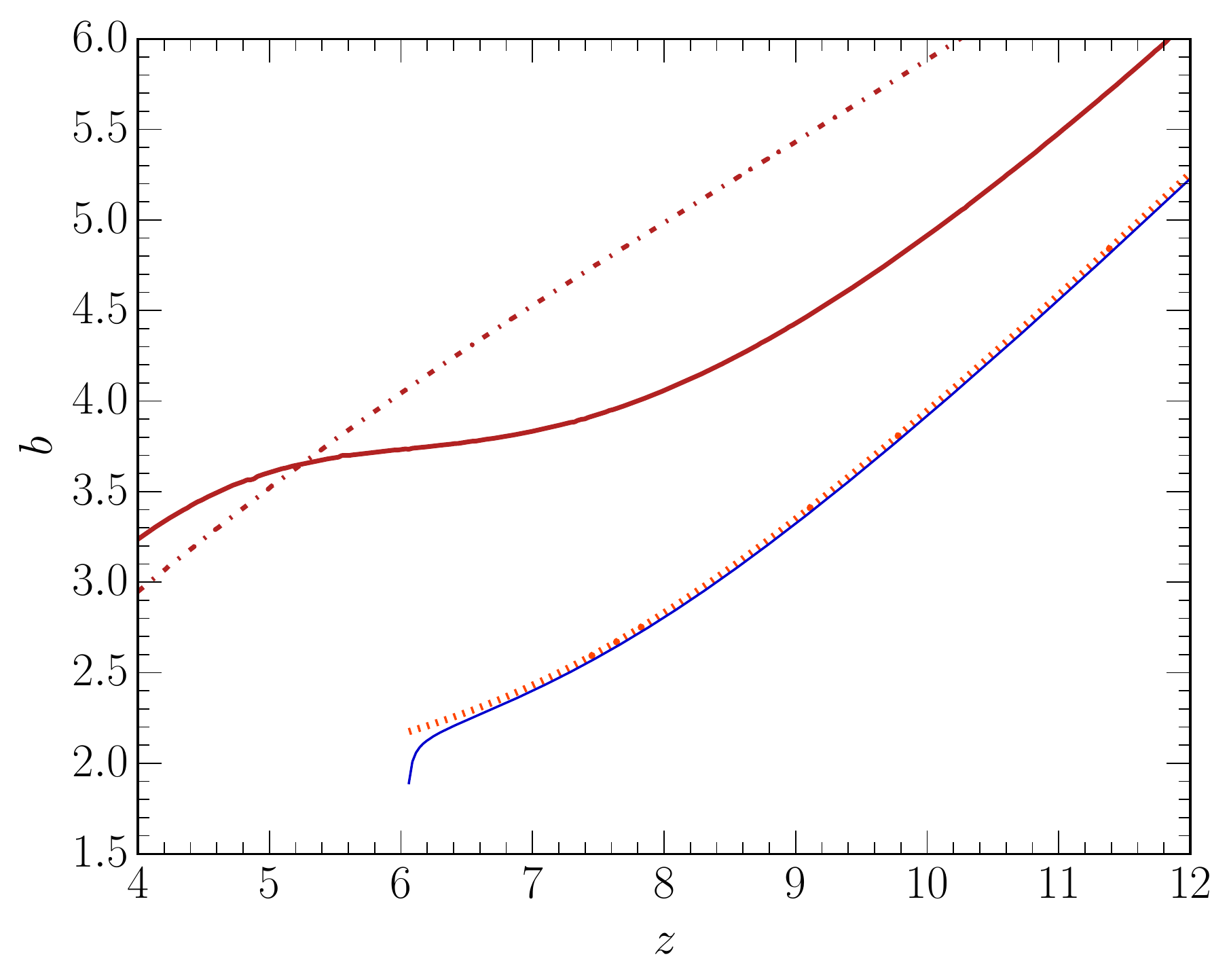}
	}
	\caption{Bias of the fluctuations in the filling factor used in eq. \eqref{psrion}.
			We show $b_Q$ (orange dotted) and $b_Qe^{-k^2 R_b^2/2}$ (blue solid) with $k = 1\, \mathrm{Mpc}^{-1}$.
			We show the mean galaxy bias $\langle b \rangle_{\alpha}$ (red dot-dashed) from \protect\cite{2001MNRAS.323....1S} weighted with the Ly$\alpha$ luminosity (eq. \eqref{meanbias}) and the mean bias weighted with the number of ionizing photons emscaping into the IGM $\langle b \rangle_{912}$ (red solid).}
	\vspace{-1\baselineskip}
	\label{fig:biasfil}
	\end{figure}

We will first analyze fluctuations in the hydrogen density. In the linear regime $\delta n_H^2 \approx 2 \delta n_H$ and:
\begin{equation}
\label{fluch2}
\delta I^\lya_\mathrm{IGM}(z, \mathbf{x}) \approx  2\delta(\mathbf{x}) I_\mathrm{IGM} 
\end{equation}	
Therefore the power spectrum is simply:
\begin{equation}
\label{psrh}
P^\lya_\mathrm{IGM,\delta\delta}(k,z) = 4(I^\lya_\mathrm{IGM})^2 \sigma^2_k(z).
\end{equation}
The treatment of fluctuations in the ionized fraction is more complex and has been discussed by \cite{2003ApJ...598..756S}; we adapt that method to our purposes. For overdensity $\delta$, we can write $\delta Q(z, \delta) \approx Q(z)(1+\delta_Q)$; if further $\delta \ll 1$, then $\delta_Q \propto \delta$. Therefore, we can define the porosity bias as
\begin{equation}
b_Q(z) \equiv \frac{\delta_Q}{\delta} = \frac{Q(z, \delta)}{Q(z)} - 1 ;
\end{equation}
To compute $Q(z, \delta)$ we use the theory developed in Sec. \ref{ssec:reion} allowing for an explicit dependence on $\delta$ in eq. \eqref{eqreion}:
\begin{multline}
\frac{dQ(z, \delta(z))}{dt} = \frac{(1 + \langle b(z) \rangle_{912}\delta(z))}{n_H(0)(1 + \delta(z))}\frac{d {Q(z, \delta = 0)}}{d{t}} - \\
- \alpha_B C(z) n_H(0)(1+\delta(z)) (1+z)^3 Q(z, \delta(z)),
\end{multline}
where $\langle b(z) \rangle_{912}$ is the mean bias weighted with the number of ionizing photons emitted by the halos (Fig. \ref{fig:biasfil}) and $\delta(z) = \delta(z=0) D(z)$, where $D(z)$ is the growth factor \citep{1996MNRAS.282..281L}.	

Finally, we have to account of the fact that the relation between $\delta_Q$ and $\delta$ is not linear on all scales: if ionized bubbles have a typical radius $R_b$, fluctuations on smaller scales must be damped. For this reason we add a factor $e^{-k^2 R_b(z)^2}$ to our final power spectrum. Our formalism is not suited to compute the typical radius of the ionized regions; thus we will use the same crude model of \cite{2003ApJ...598..756S}:
\begin{equation}
R_b = \left( \frac{1}{1-Q} \right)^{1/3} R_0,
\end{equation}
where $R_0 = 100$ kpc is a guess for the size of the ionized bubbles at the beginning of reionization. The final results are not strongly dependent on the detailed form of $R_b$ because we expect that when the bubbles reach a radius of interest in this work (e.g. of the order of tens of Mpc) the ionization fraction is almost unity and therefore this source of fluctuations is negligible. Moreover this transition is very fast and relevant only in a limited redshift range.

Fig. \ref{fig:biasfil} shows the porosity bias $b_Q$ as a function of redshift along with $b_Qe^{-k^2 R_b^2/2}$ ($k = 1 \mathrm{Mpc}^{-1}$). It is evident that the finite bubble size correction is negligible for $k < h\,\mathrm{Mpc}^{-1}$. In addition, note that  $b_Q$ is always much smaller than the average galaxy bias and that the \Lya intensity is fainter. Therefore the power spectrum of IGM recombinations is negligible when compared to the other \Lya sources.

The final power spectrum arising from fluctuations in the ionization field is:
\begin{equation}
\label{psrion}	
P^\lya_\mathrm{IGM,x_{\rm HI}}(k,z) = (I^\lya_\mathrm{IGM})^2 b_Q(z)^2 e^{-k^2R_b(z)^2} \sigma_k^2(z);
\end{equation}
In this case it is not necessary to include the shot-noise term, because fluctuations in the IGM density are not associated to discrete sources and our model is not well suited for the calculation of the shot noise arising from the discreteness of the ionized bubbles. Moreover, this source of \Lya photons is sub-dominant and shot-noise is never relevant on the scales of interest for intensity mapping.

\subsection{Fluctuations from IGM UV excitations}
\label{ssec:fluctcont}
Fluctuations associated with UV photon excitations are seeded by various contributions. First we consider the biased fluctuations in the UV emissivity:
\begin{multline}
\label{Ix}
\delta I^\lya_\mathrm{cont}(z, \mathbf{x}) = \frac{1}{4\pi} \int d\Omega \int_z^{+\infty} dz' A(z, z') \dot n_\nu(\nu', z')\times \\
\times[\langle b(z')\rangle_{\nu'} \delta(z', \mathbf{x}+ l(z, z')\hat{\Omega})],
\end{multline}
where $l(z, z')$ is the comoving distance traveled by a photon between redshift $z$ and $z'$; $A(z, z')$ is a complex function that can be recovered from eq. \eqref{Icont}, and $\langle b(z') \rangle_{\nu'}$ is the mean bias weighted with the luminosity at the appropriate UV frequency
\begin{equation}
\langle b(z')\rangle_{\nu'} = \frac{\int_{M_{\rm min}}^{+\infty} \frac{d{n}}{d{M}} L_{\nu'}(M, z) f_{\rm dust}^{\rm UV}(M, \nu', z') b(M, z') dM    }{h_P\nu\dot n_\nu(\nu', z')}.
\end{equation}
The integrand in eq. \eqref{Ix} has a non-trivial spatial dependence that tends to suppress small scale fluctuations. Indeed, the bulk of the photons that redshift into the \Lya line are emitted between the \Lya and the \Lyb line; but the distance $\l(z, z')$ required for a \Lyb photon to redshift into a \Lya photon can be quite large ($\sim 575(460)$Mpc at $z = 4 (7)$). Therefore if we consider fluctuatons on smaller scales, part of the UV photons contribute less to the power spectrum because their fluctuations do not correlate. This effect is present also for shot noise fluctuations, where fluctuations are always suppressed.

The Fourier transform of eq. \eqref{Ix} is
\begin{multline}
\delta \tilde I^\lya_\mathrm{cont}(z, \mathbf{k}) = \frac{1}{4\pi} \int d\Omega \int_z^{+\infty} dz' \times \\
\times A(z, z')\dot n_\nu(\nu', z') \langle b(z')\rangle_{\nu'} D(z') e^{i {\bf k} \cdot \hat{\Omega}l(z, z')} \tilde \delta(z = 0, \mathbf{k}) = \\
= \int_z^{+\infty} dz' A'\dot n_\nu' \langle b'\rangle D' \frac{\sin(kl')}{kl'} \tilde \delta(z = 0, \mathbf{k});
\end{multline}
where in the last equation we avoided to write explicitly the dependence in $z$ and $z'$.
Any evolution of $z$ with ${\bf x}$ was neglected; this is reasonable because on the scales relevant for intensity mapping ($10-100$Mpc) the redshift range sampled is small.

The resulting power spectrum is
\begin{multline}
\label{pscdd}
P^\lya_\mathrm{c,\delta\delta}(z, \mathbf{k}) = \left[ \int dz' A(z, z')\langle b(z') \rangle_{\nu'}\times \right. \\
\left. \times  \dot n_\nu(\nu', z') D(z') \frac{\sin(k l(z, z')}{k l(z, z')}\right]^2 \sigma^2_k(z=0).
\end{multline}	

As we will see in Sec. \ref{ssec:flucttot}, the IGM emission is very smooth on scales smaller than hundreds of Mpc; however it is significant if we consider larger scales, comparable to the typical redshift distance between the \Lya and the \Lyb line. For this reason we neglected shot-noise, because the discrete nature of the sources is not important on such large scales.

Other fluctuations can arise in the Gunn-Peterson optical depth, through the \HI density. It affects both the absorption probability outside the integral \eqref{Icont} and the transmission fractions. These fluctuations are suppressed because, as discussed in Sec. \ref{ssec:meancont}, we can not consider photons absorbed in the \Lya line since at first order the UV intensity field is homogeneous. The bulk of the \Lya photons emitted in the IGM originates from the absorption of photons emitted between the \Lya and the \Lyb line, therefore the mean \Lya intensity from the excitations of other Lyman lines is dominated by the ISM emission by orders of magnitude.

\subsection{Total power spectrum}
\label{ssec:flucttot}
In this Section we will examine the results for the total Ly$\alpha$ intensity power spectrum predicted by our model. For simplicity we have neglected the correlations among different processes and simply summed the individual terms. This procedure, although approximate, yields a conservative estimate, as such cross-correlations might increase the power spectrum amplitude up to a factor of 2 (since  $\frac{(a+b)^2}{a^2 + b^2} = 2 - \frac{(a-b)^2}{a^2 + b^2}$).

\begin{table}
\centering
\begin{tabular}{cccc}
\toprule
 & $k = 0.01$  & $0.1$ & $1$ \\
\midrule

$z=4$ & $4.72\times 10^{-6}$ & $9.76\times 10^{-4}$ & $1.71\times 10^{-2}$ \\
$6$ & $5.50\times 10^{-7}$ & $6.50\times 10^{-5}$ & $1.15\times 10^{-3}$ \\
$7$ &  $2.00\times 10^{-7}$ & $2.09\times 10^{-5}$ & $3.63\times 10^{-4}$ \\
$8$ &  $7.67\times 10^{-8}$ & $7.52\times 10^{-6}$ & $1.27\times 10^{-4}$ \\
$10$ &  $1.35\times 10^{-8}$ & $1.19\times 10^{-6}$ & $1.92\times 10^{-5}$ \\
\bottomrule
\end{tabular}
\caption{Predicted \Lya PS at different redshifts $z$ and wavelengths $k$. We show $k^3 P^\lya(k, z)/2\pi^2$ in $[{\rm nW}^2{\rm m}^{-4}{\rm sr}^{-2}]$, while $k$ is in $[h {\rm Mpc}^{-1}]$.}
\label{tab:psvalue}	
\end{table}

\begin{figure}
		\vspace{+0\baselineskip}
		{
		\includegraphics[width=0.45\textwidth]{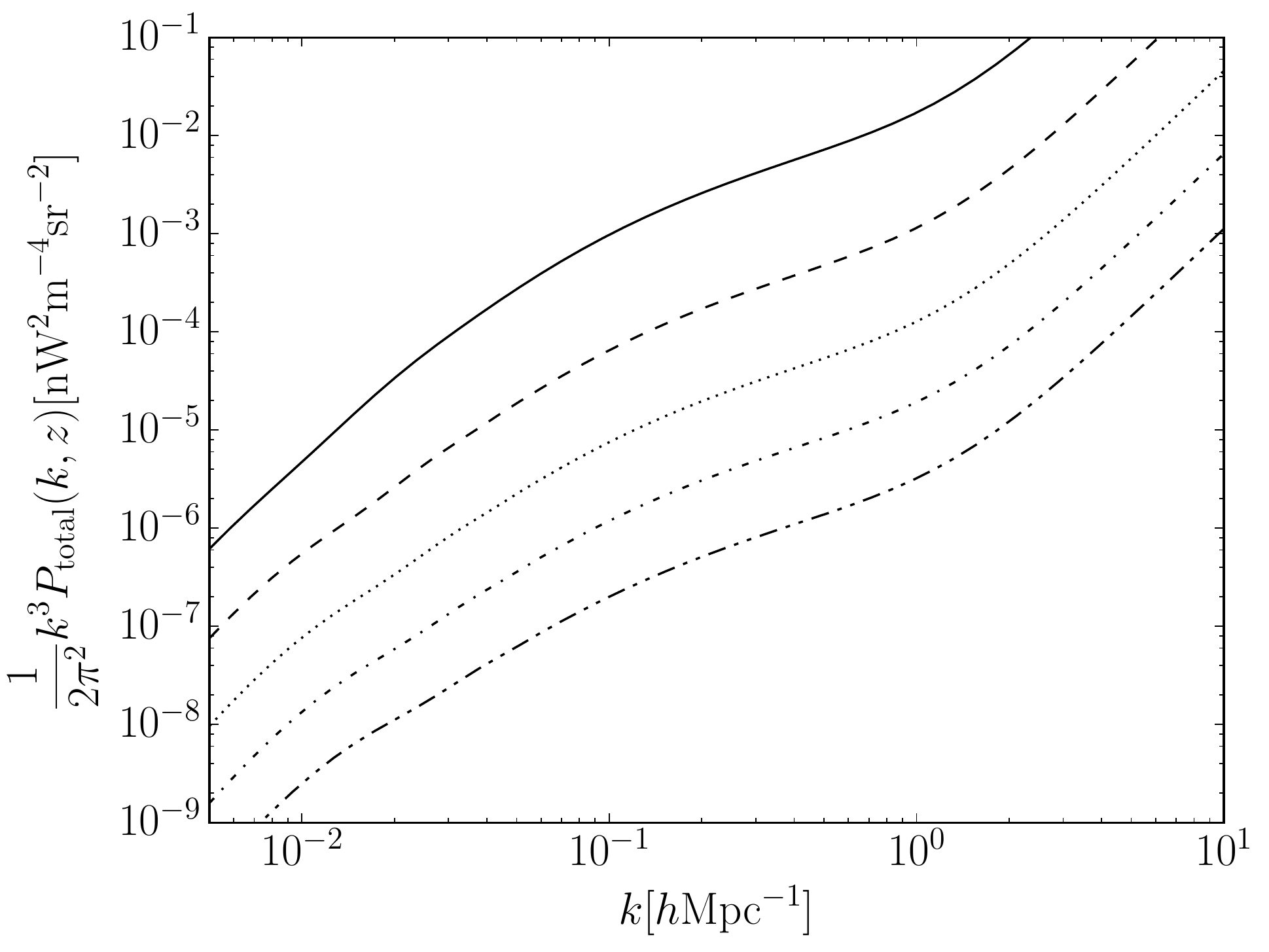}
		}
		\caption{Total Ly$\alpha$ power spectrum at $z = ~4,~6,~8,~10,~12$ (from top to bottom).}
		\vspace{-1\baselineskip}
		\label{fig:psinttot}
	\end{figure}

\begin{figure}
		\vspace{+0\baselineskip}
		{
		\includegraphics[width=0.45\textwidth]{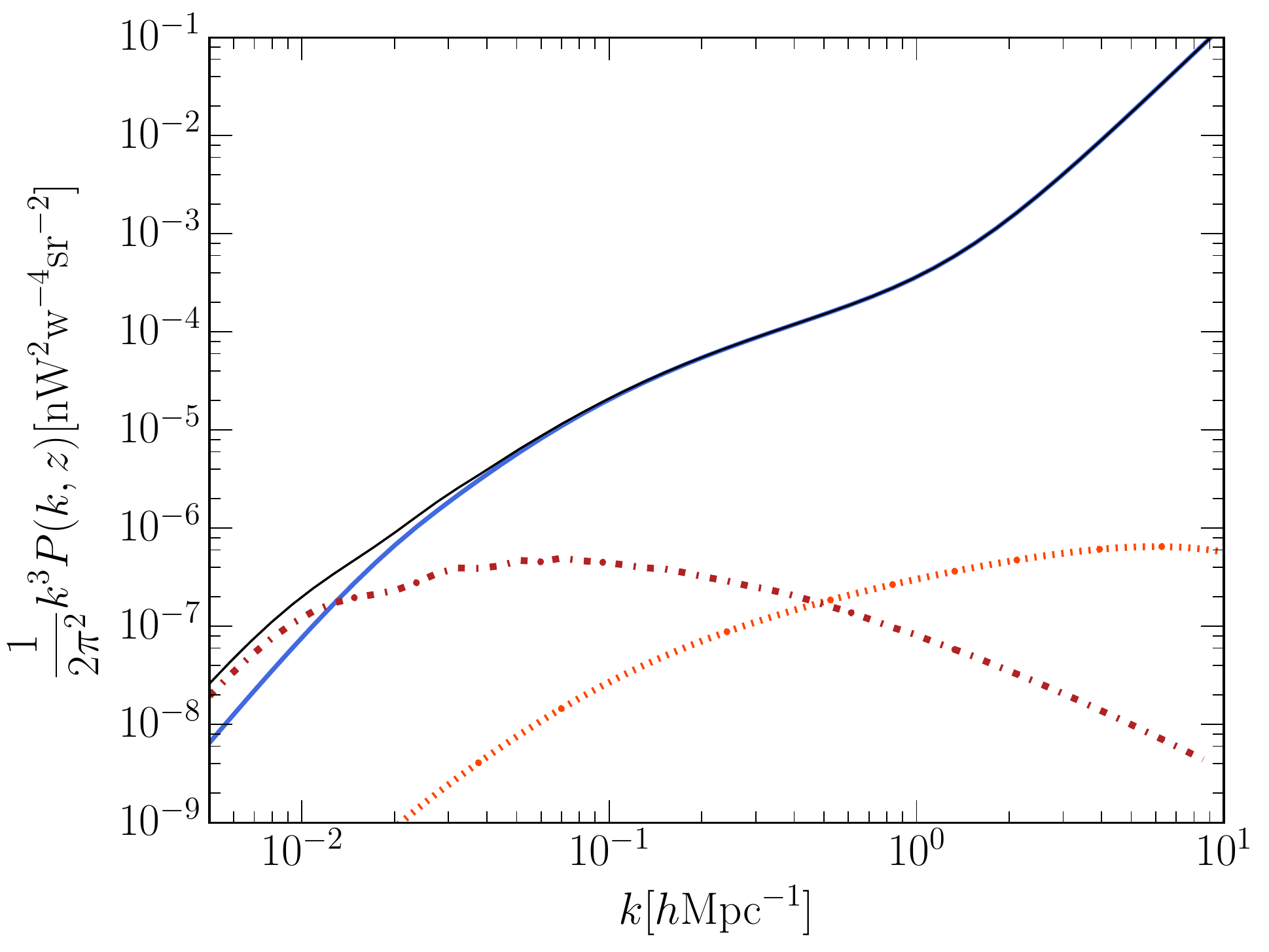}
		}
		\caption{Decomposition of the power spectrum in the different processes at $z=7$.
			We show the diffuse emission in the IGM from UV excitations (red dot-dashed), the emission from the ISM in the halos (blue solid); the recombinations in the IGM (yellow dotted) and the total PS (black solid).}
		\vspace{-1\baselineskip}
		\label{fig:psinttotcomp7}
	\end{figure}

\begin{figure}
		\vspace{+0\baselineskip}
		{
		\includegraphics[width=0.45\textwidth]{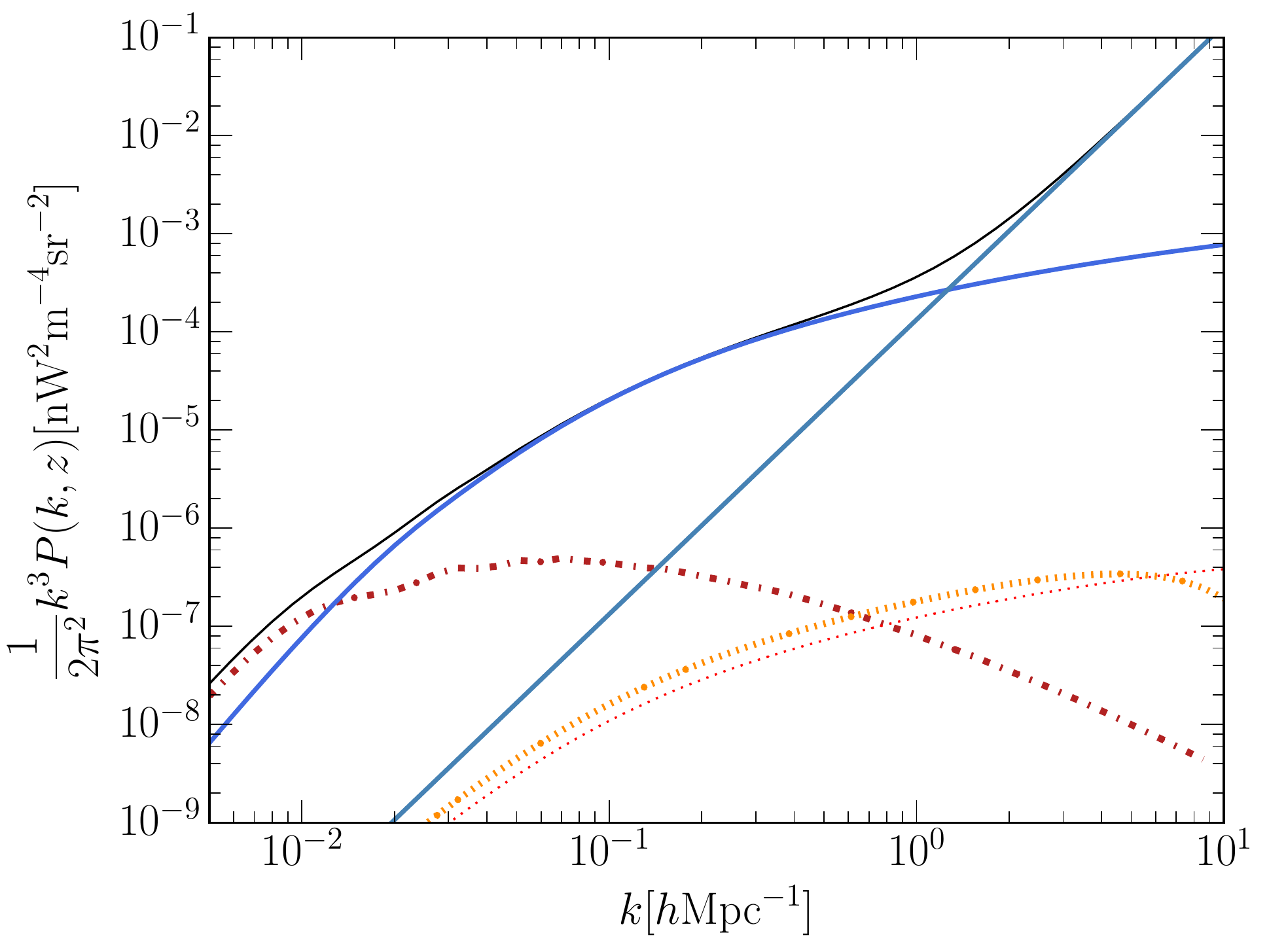}
		}
		\caption{Contribution to the \Lya PS from the different processes considered at $z=7$. We show total PS (black solid); the linear fluctuations (blue solid) and the shot-noise (light-blue solid) from ISM emission (eq. \eqref{pshlin} and \eqref{pshsn}); from IGM emission from UV excitations (red dot-dashed, eq. \eqref{pscdd}); from recombinations in the IGM (dotted lines), from fluctuations in the filling factor (thin red, eq. \eqref{psrion}) and in the density (thick yellow, eq. \eqref{psrh}). 
		}
\vspace{-1\baselineskip}
		\label{fig:psintcontcomp7}
	\end{figure}

Table \ref{tab:psvalue} shows the numerical values of the total \Lya PS, $P^\lya(k, z)$, predicted by our model and Fig. \ref{fig:psinttot} shows its redshift evolution.
As expected the power monotonically decreases with redshift, following the analogous mean \Lya intensity trend. Since this is the main observable for proposed intensity mappers (\cite{2013ApJ...763..132S, 2014ApJ...786..111P, 2014arXiv1412.4872D}) with a signal-to-noise too low to perform tomographic studies, it is important to understand what we can learn from such measurement. The power spectrum shape shows two key features: a pronounced change of slope at $k \approx 1 h {\rm Mpc}^{-1}$ and, for $z \gsim 6$, a bump at $k \approx 0.01 h {\rm Mpc}^{-1}$. As we will se from Fig. \ref{fig:psinttotcomp7} and Fig. \ref{fig:psintcontcomp7}, they describe changes in the well defined physical properties: shot-noise on small scales and IGM emission on large scales. 

In Fig. \ref{fig:psinttotcomp7} we show the contributions to the power spectrum from the three processes considered at $z=7$. The ISM emission dominates the power spectrum by orders of magnitude at $k \gsim 0.1 h{\rm Mpc}^{-1}$, but on larger scales the IGM emission from UV excitations can be significant. The exact $k$ in which the PS starts to probe the IGM emission can depend on the choices made in our model; however it is important to understand that IGM emission should be taken into account when considering \Lya intensity fluctuations on large scales and that we can discriminate from an observation if we are observing the ISM or the IGM emission from the slope of the PS. It is also interesting to notice that even if in our model $\sim 70\%$ of the \Lya photons are emitted by the IGM, their distribution is very smooth and therefore they do not give a significant contribution to the PS in all but the largest scales. 

Fig. \ref{fig:psintcontcomp7} shows the PS of the different fluctuations considered at $z=7$. At $k < 1 h {\rm Mpc}^{-1}$, the PS is always dominated by linear fluctuations in the \Lya or UV emission from the galaxies, therefore even the IGM emission traces the SFR distribution. Recombinations in the ionized IGM are always subdominant, for the low IGM density. Shot noise is significant only for $k > 1 h {\rm Mpc}^{-1}$, but on these scales \Lya radiative transfer effects are important, because the \Lya emission can be spread over an extended halo. Since the shot-noise power spectrum is constant over $k$, any deviation from this dependence can be used to constrain these effects.
 
\begin{figure}
		\vspace{+0\baselineskip}
		{
		\includegraphics[width=0.45\textwidth]{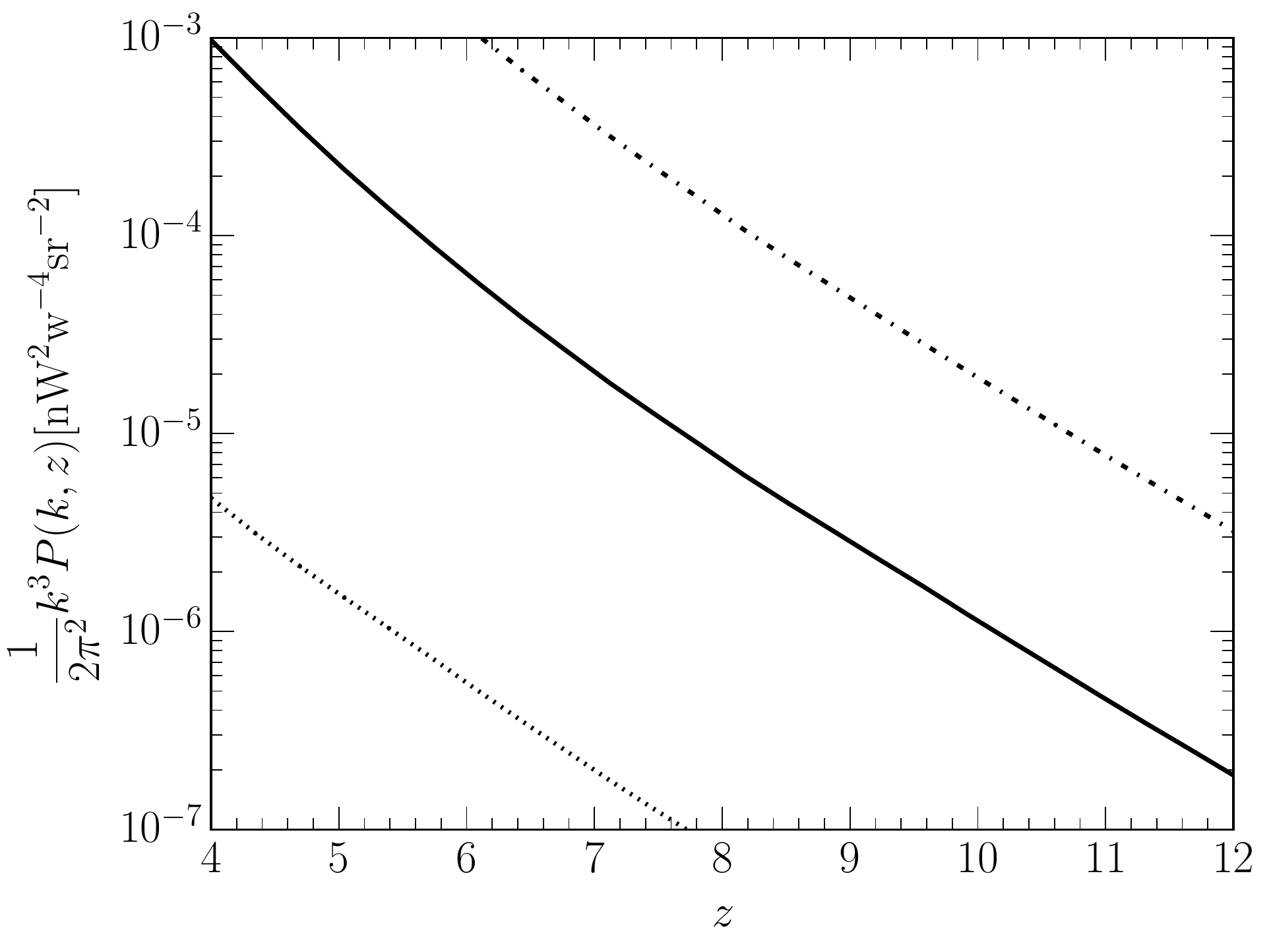}
		}
		\caption{Redshift evolution of the \Lya PS. We show $P^\lya$ for  $k = 0.1 h{\rm Mpc}^{-1}$ (solid), $k = 0.01 h {\rm Mpc}^{-1}$ (dotted) and $k = 1 h{\rm Mpc}^{-1}$ (dot-dashed).
		}
\vspace{-1\baselineskip}
		\label{fig:psredevol}
	\end{figure}

Fig. \ref{fig:psinttotcomp7} and \ref{fig:psintcontcomp7} refer only to the $z = 7$ case; at different redshifts the qualitative results are similar, while the quantitative change can be inferred from Fig. \ref{fig:psinttot}. The only difference is that at $z < 6$ the IGM is not completely opaque to \Lya photons and therefore the IGM emission from UV excitations is less prominent compared to the ISM one even at large scales. 
Finally, in Fig. \ref{fig:psredevol} we show the redshift evolution of the PS for $k = 0.1 h {\rm Mpc}^{-1}$. The dominant feature is a smooth decline, i.e. a distinctive, model independent signature associated with, e.g., reionization is not visible. This is expected since \Lya intensity mapping is a better tracer of SFR history rather than of the IGM thermal/ionization state.     

It should be stressed that the power spectrum scales as the square of the intensity: for example a factor $3$ in the \Lya escape fraction translates to an order of magnitude in the power spectrum. Therefore the quantitative results from our model depend strongly on the choices made for the poorly constrained parameters that we introduced. However the uncertainty on the numerical value of our results does not prevent a clear understanding of the physics involved in the \Lya PS; moreover, the use of an analytical model allows a precise control of the dependence of the results on free-parameters.


\section{Summary and conclusions}
\label{sec:conc}
We have developed an analytical model to estimate the \Lya emission at $z > 4$. Our goal is to predict the power spectrum of the spatial fluctuations that could be observed by an intensity mappig survey. In particular we aim at understanding what physical processes occurring in the EoR can be probed with this technique. Our model uses the latest data from the HST legacy fields and the abundance matching technique to associate UV emission and dust properties with the halos, computing the ISM and IGM emission consistently.

The \Lya intensity from the diffuse IGM emission is 1.3 (2.0) times more intense than the ISM emission at $z = 4(7)$; both components are fair tracers of the star-forming galaxy distribution. However the power spectrum is dominated by ISM emission on small scales  ($k > 0.01 h{\rm Mpc}^{-1}$) with shot noise being significant only above $k = 1 h{\rm Mpc}^{-1}$. At very lange scales ($k < 0.01h{\rm Mpc}^{-1}$) diffuse IGM emission becomes important. The comoving \Lya luminosity density from IGM and galaxies, $\dot \rho_\lya^{\rm IGM}  = 8.73(6.51) \times 10^{40} {\rm erg~}{\rm s}^{-1}{\rm Mpc}^{-3}$ and $\dot \rho_\lya^{\rm ISM}  = 6.62(3.21) \times 10^{40} {\rm erg~}{\rm s}^{-1}{\rm Mpc}^{-3}$ at $z = 4(7)$, is consistent with recent SDSS determinations.
We predict a power $k^3 P^\lya(k, z)/2\pi^2 = 9.76\times 10^{-4}(2.09\times 10^{-5}){\rm nW}^2{\rm m}^{-4}{\rm sr}^{-2}$ at $z = 4(7)$ for $k = 0.1 h {\rm Mpc}^{-1}$. 

The quantitative results from our model depend on the choice of the free-parameters. However, at least three points appear to solidly emerge from the analysis: (i) \Lya intensity mapping is a good probe of LSS and of star formation at high redshift; (ii) both the IGM and the ISM produce a significant, and potentially detectable, \Lya emission, with the IGM one being is smoother and becoming important only on $\simgt 100$ Mpc scales; (iii) the shape of the power spectrum is important to constrain the physics of \Lya emission and to understand the nature of the sources.   

Whether an intensity mapping survey is observationally feasible is still unclear at this time. Even if the outlook for the HI 21 cm line (the first intensity mapping survey proposed) seems positive and several instruments are already producing data or are under construction, other lines present completely different challenges (such as line confusion). Therefore further study is necessary, in particular to test the extraction of the predicted power spectrum when actual foregrounds are included. Moreover, modeling  the physics involved in the line emission process is an extremely challenging task. For this reason,  the \Lya line emerging from galaxies should be carefully calibrated against observations. In spite of these difficulties, a successful intensity mapping survey would produce such a breakthrough that any effort towards this goal is certainly worthwhile. In a forthcoming paper we will build upon the present results and assess the feasibility of a \Lya intensity mapping experiment by presenting possible strategies to control foreground(s) removal. 

\section*{Acknowledgments}
We thank S. Gallerani, B. Yue, A. Mesinger and E. Sobacchi for helpful discussions.

\bibliography{paper1}

\end{document}